\documentclass[10pt,aps,prc,amsmath,amssymb,twocolumn,floatfix,nofootinbib,showpacs,superscriptaddress]{revtex4-1}
\usepackage[utf8]{inputenc}

\usepackage{amssymb,amsthm,amsmath,amstext,amsbsy}
\usepackage{dcolumn}
\usepackage{bbm}
\usepackage{url}
\usepackage{nicefrac}
\usepackage{graphicx}
\usepackage{hyperref}
\usepackage{verbatim}
\usepackage[usenames,dvipsnames]{xcolor}
\usepackage{marginnote}
\usepackage{csquotes}
\usepackage{multirow}

\usepackage{booktabs}

\graphicspath{{figures/},{new_figures/}}

\newcommand{\beq}{\begin{equation}}
\newcommand{\eeq}{\end{equation}}

\newcommand{\bseq}{\begin{subequations}}
\newcommand{\eseq}{\end{subequations}}

\newcommand{\ts}{\textstyle}
\newcommand{\ds}{\displaystyle}

\newcommand{\NNLO}{\ensuremath{{\rm N}{}^2{\rm LO}}}
\newcommand{\NNNLO}{\ensuremath{{\rm N}{}^3{\rm LO}}}
\newcommand{\NNNNLO}{\ensuremath{{\rm N}{}^4{\rm LO}}}

\newcommand{\bi}{\begin{itemize}}
\newcommand{\ei}{\end{itemize}}
\newcommand{\I}{\item}
\newcommand{\be}{\begin{enumerate}}
\newcommand{\ee}{\end{enumerate}}
\newcommand{\bc}{\begin{center}}
\newcommand{\ec}{\end{center}}

\newcommand{\pr}{\text{pr}}

\newcommand{\kmax}{k_{\rm max}}

\newcommand{\cbar}{\bar{c}}
\newcommand{\cbarsubk}[1]{\cbar_{(#1)}}
\newcommand{\cbark}{\cbarsubk{k}}
\newcommand{\cbarkpone}{\cbarsubk{k+1}}
\newcommand{\cbarmin}{\cbar_{<}}
\newcommand{\cbarmax}{\cbar_{>}}

\newcommand{\pdf}{\pr}  % {f}   % adjust this as desired.  E.g., f --> \pr
\newcommand{\feps}{\pdf}   % {\pdf_{\epsilon}}
\newcommand{\CHbar}{$\overline{\rm CH}$}
\newcommand{\CH}{CH}

\newcommand{\Q}{Q}

\newcommand{\Elab}{E_{\rm lab}}

\newcommand{\signp}{\sigma_{np}}
\newcommand{\sigmaref}{\sigma_{\rm ref}}

\newcommand{\Data}{\ensuremath{c_0,\ldots,c_k}}
\newcommand{\Aeps}{\ensuremath{{\rm A}_{\epsilon}}}
\newcommand{\Ceps}{\ensuremath{{\rm C}_{\epsilon}}}
\newcommand{\Aepsone}{\Aeps^{(1)}}
\newcommand{\Cepsone}{\Ceps^{(1)}}
\newcommand{\Aone}{\ensuremath{{\rm A}^{(1)}}}
\newcommand{\Bone}{\ensuremath{{\rm B}^{(1)}}}
\newcommand{\Cone}{\ensuremath{{\rm C}^{(1)}}}
\newcommand{\Deltakone}{\ensuremath{\Delta_k^{(1)}}}
\newcommand{\dkp}{\ensuremath{d_k^{(p)}}}

\newcommand{\ppercentt}{\ensuremath{\left(p\%\right)_{\rm t}}}

\newcommand{\LamchiSB}{\Lambda_{\chi {\rm SB}}}
\newcommand{\chiPT}{\chi{\rm PT}}

\newcommand*{\Scale}[2][4]{\scalebox{#1}{$#2$}}%

\begin{document}

\title{Quantifying truncation errors in effective field theory}

\author{R.J.~Furnstahl}
\email{furnstahl.1@osu.edu}
\affiliation{Department of Physics, The Ohio State University, Columbus, OH 43210, USA}

\author{N.~Klco}
\email{nk405210@ohio.edu}
\author{D.R.~Phillips}
\email{phillips@phy.ohiou.edu}
\affiliation{Institute of Nuclear and Particle Physics and Department of Physics and
Astronomy, Ohio University, Athens, OH 45701, USA}

\author{S.~Wesolowski}
\email{wesolowski.14@osu.edu}
\affiliation{Department of Physics, The Ohio State University, Columbus, OH 43210, USA}
\date{\today}

\begin{abstract}
Bayesian procedures designed to quantify truncation errors in perturbative
calculations of quantum chromodynamics observables are adapted to expansions
in effective field theory (EFT).
In the Bayesian approach, such truncation errors are derived from
degree-of-belief (DOB) intervals for EFT predictions.
Computation of these intervals requires specification of prior probability
distributions (``priors") for the expansion coefficients.
By encoding expectations about the naturalness of these coefficients, this
framework provides a statistical interpretation of the standard EFT procedure where
truncation errors are estimated using the order-by-order convergence of the expansion.
It also permits exploration of the ways in which such error bars are, and are not,
sensitive to assumptions about EFT-coefficient naturalness.
We first demonstrate the calculation of Bayesian probability distributions
for the EFT truncation error in some representative examples, and then
focus on the application of chiral EFT to neutron-proton scattering. Epelbaum, Krebs, and
Mei{\ss}ner recently articulated explicit rules for estimating truncation errors in such
EFT calculations of few-nucleon-system properties.
We find that their basic procedure emerges generically from one class of
naturalness priors considered, and that all such priors result in consistent quantitative
predictions for 68\% DOB intervals.
We then explore
several methods by which the convergence properties of the EFT for a set of
observables may be used to check the statistical consistency of the EFT expansion parameter.
\end{abstract}

\smallskip

\maketitle

%%%%%%%%%%%%%%%%%%%%%%%%%%%%%%%%%%%%%%%%%%%%%%%%%%%%%%%%%%%%%%

\section{Introduction} \label{sec:intro}

Effective field theories (EFTs) describe the physics of systems with a separation of scales\footnote{Pedagogical introductions to EFTs can be found in
Refs.~\cite{Kaplan:1995uv,Phillips:2002da,Epelbaum:2010nr}.}.
A key element in any EFT is a power counting that organizes calculations
into an expansion in a dimensionless parameter or parameters, which are typically formed
from ratios of the low-energy and high-energy scales in the system under consideration.
We denote this parameter generically as $\Q$. In the simplest case $\Q=p/\Lambda_b$ is the ratio of the typical momentum, $p$,
of the process of interest to the break-down scale, $\Lambda_b$, of the EFT---i.e., the scale at which the first dynamics not explicitly
included in the EFT appears.
Even in more complex situations with many low-energy scales, the EFT expansion for $X$ can be denoted:
\begin{equation}
X=X_0 \sum_{n=0}^{\infty} c_n \Q^n  \;,
\label{eq:obsexp}
\end{equation}
where $X_0$ is the natural size of the observable $X$, and $\{c_n\}$ are dimensionless coefficients, some of which may be zero.
In most EFTs the expansion (\ref{eq:obsexp}) is inherited
directly from the EFT Lagrangian or potential---with suitable additions (e.g., terms
of the form $Q^n \log(Q)$) due to quantum corrections. In nuclear physics, however,
the dynamics is intrinsically non-perturbative, and there exists at least some sub-class of EFT graphs that
must be summed to all orders. The connection between
the Lagrangian and observables is then less direct. Nevertheless, a properly formulated EFT for
nuclear physics is expected to have a $Q$-expansion for observables of the form (\ref{eq:obsexp}) and
it is the properties of such expansions which are our concern in this paper.

A key benefit of the perturbative series (\ref{eq:obsexp}) is that it permits estimation of the error induced
by truncation at a finite order $k$: ``truncation errors". If the coefficients $c_n$ for an observable were
to vary significantly and unsystematically in size, the expansion (\ref{eq:obsexp})
would be unsuited to this end. However, experience, and the principle
of naturalness, suggest that the coefficients \emph{are}
typically of order one---even in the more complex nuclear context.

In Ref.~\cite{Furnstahl:2014jpg} we laid out a recipe for uncertainty quantification in
EFTs for nuclear physics. While they are not the only source of theory uncertainty,
truncation errors are often the dominant uncertainty in EFT calculations. 
We argued that Bayesian methods~\cite{Schindler:2008fh}
provide an error bar,
 with a well-founded
statistical interpretation,
that accounts for {\it all} sources of uncertainty in the EFT.
In particular, Bayesian methods are essential to the assessment of truncation error: assumptions
(or expectations) about the EFT are encoded
in ``prior probability distribution functions" (pdfs).
The Bayesian approach then proceeds by integrating out (``marginalizing'')
the coefficients of omitted terms to establish the truncation error.

The use of priors is often controversial because they can introduce
subjective judgments about, e.g., the meaning of naturalness, into
the computation. We argue that, on the contrary, introducing and stating
Bayesian priors on higher-order EFT coefficients
renders explicit in the calculation assumptions that are present but typically not articulated. This
allows such assumptions to be applied consistently, tested, and
modified in light of new information.

In this work we begin with the general formalism for computation of truncation errors discussed
in the context of perturbative QCD (pQCD) by Cacciari and
Houdeau in Ref.~\cite{Cacciari:2011ze} and further developed in
Refs.~\cite{Bagnaschi:2014wea,Bagnaschi:2014eva}, where it is called
\CHbar\ (cf.~Ref.~\cite{Bagnaschi:2015iea} for a brief summary). 
We rederive, adapt, and extend their prescription to EFT
expansions. We explore several different choices of prior for the coefficients
$\{c_n\}$ and examine---within some generic examples---how such prior choice affects the truncation-error estimate.
We then look specifically at the nuclear context, focusing on the extent to which such calculations justify
the uncertainty quantification (UQ) procedure typically adopted in EFTs. This procedure has recently been clearly stated and applied to nucleon-nucleon (NN) cross sections
by Epelbaum, Krebs,
and Mei{\ss}ner (henceforth EKM) in their fourth- and fifth-order applications of chiral EFT to these observables~\cite{Epelbaum:2014efa,Epelbaum:2014sza}.
(Introductions to chiral EFT can be found in Refs.~\cite{Weinberg:1978kz,Gasser:1983yg,Jenkins:1990jv,Bernard:2006gx,Bedaque:2002mn,Epelbaum:2005pn}.)
Note that here we do not deal with the extent to which truncation errors affect the low-energy constants (LECs) extracted from fitting
EFT expansions to data. This will be discussed in a separate publication~\cite{Wesolowski:2015}. Our focus here is solely on
estimating the truncation error at order $k$ for the series (\ref{eq:obsexp}), given the assumption of naturalness, and information on the
size of the coefficients $c_0, \ldots, c_k$.

In Section~\ref{sec:CH}, we provide a brief overview of the Bayesian rules
we will need and then adapt the \CH\ prescription so that it is suitable
for application to EFT expansions.
We also enlarge the set of priors considered by CH.
In Section~\ref{sec:EKM}, this approach is applied to some two-body
observables considered by EKM, using their assumed breakdown scale to compare to their
error assessment. In Section~\ref{sec:scalechecking} we
explore methods to determine the breakdown scale
from the requirement that the EFT coefficients be consistent with naturalness.
In Section~\ref{sec:conclusion} we summarize our results.

%%%%%%%%%%%%%%%%%%%%%%%%%%%%%%%%%%%%%%%%%%%%%%%%%%%%%%%%%%%%%%

\section{Adapting \CH\ to EFT} \label{sec:CH}

\subsection{Setting up the problem}

Consider the perturbative series (\ref{eq:obsexp}) for the observable $X$. If the series is truncated at order $k$ then
the error induced is $X_0 \Delta_k$, where the scaled, dimensionless parameter that determines the truncation error is:
\beq
  \Delta_k \equiv \sum_{n=k+1}^\infty c_n \Q^n
  \;,
  \label{eq:Deltak_def}
\eeq
provided the series actually converges and is not solely asymptotic. For sufficiently small values of $\Q$, the first omitted term $c_{k+1} \Q^{k+1}$ is a good estimate for $\Delta_k$.
This leads to simplified formulas for the evaluation of DOB intervals. Below we will consider both this first-term approximation and evaluations at larger $\Q$ that use several terms in $\Delta_k$. In either case this provides an estimate of the deviation of the series at order $k$ from the
true value of the observable---even if the series is asymptotic.

In Ref.~\cite{Cacciari:2011ze} Cacciari and Houdeau (hereafter ``CH") considered the case that the
series (\ref{eq:obsexp}) is a pQCD expansion. The expansion is then
in powers of the strong
coupling, $\alpha_s(\mu)$, where $\mu$ is a renormalization scale chosen
appropriately for the observable $X$. The optimal choice of $\mu$ is the
subject of much debate and many prescriptions in the literature. Indeed,
the variation of the truncated-at-order-$k$ expression for $X$ under an order-unity change
 in $\mu$ is canonically used to estimate $\Delta_k$. This is justified because of the truncated expansion's residual dependence
on the renormalization scale $\mu$; the full sum should be independent
of $\mu$, so the variation with $\mu$ contains information about
omitted terms.

CH pointed out that varying the scale $\mu$ around an optimal value $\mu_0$, say, between $\mu_0/2$ and $2 \mu_0$, does
not yield an uncertainty with a straightforward statistical interpretation. They therefore
laid out a Bayesian probability-theory calculation of $\Delta_k$. Ultimately estimates from scale variation seem to coincide quite well
with the results of this more rigorous probabilistic analysis.
As we now describe, that analysis starts with priors for $\{c_n\}$ that encode the assumption that these pQCD coefficients
are of order unity (once the typical size of the observable $X$ is
factored out of the expression). It modifies them in light of information acquired as more orders
in the series for $X$ are computed, and ultimately obtains a
posterior pdf for $\Delta_k$. With this posterior in hand, either
the degree of belief (DOB) corresponding to a given
interval of values of the truncation error, or the range of truncation errors
corresponding to a specified
DOB, can be computed.

CH's analysis of truncation errors in pQCD took the case where the series
is in powers of $\alpha_s$, rather than $Q$. Later work~\cite{Bagnaschi:2014wea,Bagnaschi:2014eva} introduced a scale factor $\lambda$, such that the
expansion was in powers of $\alpha_s/\lambda$ (e.g., because the expansion parameter might really be $\alpha_s/\pi$
or include a color factor) and a possible combinatoric
factor (such as $n!$ from high-order renormalon contributions).
In this, termed the ``\CHbar\ prescription", the assumption is modified to say that the coefficients in the perturbative series of an
appropriately chosen expansion parameter are distributed such that they
share a common upper bound. In EFT the rescaling by a factor $\lambda$ corresponds to the choice of a different breakdown scale for the EFT, and we discuss
this possibility in Section~\ref{sec:scalechecking}. We do not consider the combinatoric factor,
which has not been identified in EFT expansions for few-nucleon-system observables.

Since one of the low-momentum scales of which $Q$ is formed is the momentum at which the observable $X$ is measured,
the EFT expansion parameter is strongly dependent on kinematics. While the QCD expansion's convergence can be improved by matching the scale at which $\alpha_s$ is evaluated to that of the observable of interest, in EFT the dependence of $\Q$ on momentum is intrinsic---not a matter of choice. Furthermore, the  high-momentum scale, $\Lambda_b$, that specifies the radius of convergence of the EFT momentum expansion, may not be known {\it a priori}, it may only be able to be inferred from the behavior of the EFT series. This is a key difference between EFT and pQCD, since in pQCD, the value of $\alpha_s$ can always be specified. In EFT a value of $\Q$ corresponding to a particular momentum must be chosen, and then checked for consistency. Complicating the choice of an appropriate $\Lambda_b$---and concomitantly the evaluation of $\Q$---for many low-energy EFTs is that at least some of the $c_n$'s need to be extracted from data, either from $X$ itself or from other observables.

These differences from the pQCD situation are reflected in the need to check the naturalness of EFT coefficients for a given choice of expansion parameter, something that we discuss in detail in Section~\ref{sec:scalechecking}.
For the time being we assume that $\Lambda_b$ has been determined as part of the steps in the EFT analysis that yielded the coefficients $\{c_n: n=0, \ldots, k\}$. Empirically EKM found that
$\Lambda_b \approx 600$ MeV resulted in natural coefficients in their EFT series for neutron-proton scattering cross sections~\cite{Epelbaum:2014efa,Epelbaum:2014sza}. However,
the non-perturbative nature of NN scattering makes it unclear how naturalness for
EFT LECs results in these apparently natural values of $\{c_n: n=0, \ldots, k\}$ in the cross-section's expansion.
This connection is very clear for perturbative EFT expansions of interest in nuclear physics, e.g.,
the chiral expansion for the nucleon mass
(see Ref.~\cite{Schindler:2008fh}) and the expansion for energy per particle
of a dilute Fermi
system with natural-sized scattering length~\cite{Hammer:2000xg}. Regardless though, in either perturbative or non-perturbative cases, an incorrect choice of high-momentum scale, $\Lambda_b$, will result in coefficients that are {\it not} natural. This emphasizes the close connection between the assumption of a particular expansion parameter and the imposition of a naturalness prior.

\subsection{Conditional probabilities: definitions and rules}

We use the notation $\pdf(x|I)$ to denote the probability (density) of
$x$ given information $I$; thus $\pdf(\Delta_k|c_0, \ldots, c_k)$ is the
desired pdf for $\Delta_k$.
%In the following, we assume that $\pdf(x)$ is always normalized:
%$\int\!dx\, \pdf(x) = 1$, where the integration is over the appropriate range
%of $x$.
The specification
$c_0, \ldots, c_k$ suggests that $c_0$ is non-zero, but it is straightforward to generalize the results derived
here to the case where the first non-zero coefficient is $c_l$ with $l>0$ (as often is the
case in QCD) or that where some
intermediate coefficients are identically zero (as for chiral EFT in NN scattering
where $n=1$ does not appear).

Because the terminology, techniques, and manipulations of Bayesian statistics
may be unfamiliar to our intended audience, we include a brief overview
here of those aspects needed for the \CH\ procedures~\cite{Sivia:06,Gregory:05}.
We indicate parenthetically some analogies to familiar manipulations in quantum
mechanics.
We emphasize that the correspondences are not to be taken literally.

Bayesian probabilistic inference is built on the sum and product rules.
If the set $\{x_i\}$ is exhaustive
and exclusive (cf.\ complete and orthogonal), then the sum rule says
that $\pdf(x_i|I)$ is normalized,
\beq
  \sum_i \pdf(x_i | I) = 1
     \quad \longrightarrow \quad
   \int\!dx\, \pdf(x|I) = 1
   \;,
   \label{eq:sum}
\eeq
where the continuum version is integrated over the appropriate range of $x$.
But it further implies marginalization (cf.\ inserting a complete
set of orthonormal basis states):
\beq
  \pdf(x|I) = \sum_j \pdf(x,y_j|I)
  \;,
\eeq
or the continuum version
\beq
   \pdf(x|I) = \int\!dy\, \pdf(x,y|I) = 1
   \;,
   \label{eq:marginalize}
\eeq
where $\pdf(x,y|I)$ is the joint probability of $x$ and $y$ given $I$.
We will apply this repeatedly, either to introduce new parameters or to
integrate out ``nuisance'' parameters.

Expressing $\pdf(x|I)$ in terms of the joint probability $\pdf(x,y|I)$
through Eq.~\eqref{eq:marginalize}
leads to progress by applying the product rule to relate it
to other pdfs:
\beq
  \pdf(x,y|I) = \pdf(x|y,I)\,\pdf(y|I)
              = \pdf(y|x,I)\,\pdf(x|I)
              \;.
       \label{eq:product}
\eeq
The first equality translates to: ``the joint probability of $x$ and $y$
is equal to the probability of $x$ given $y$ and $I$ times the probability
of $y$ given $I$.''
The second equality follows by symmetry, but when rearranged becomes
Bayes' theorem:
\beq
  \pdf(x|y,I) = \frac{\pdf(y|x,I)\, \pdf(x|I)}{\pdf(y|I)}
  \;,
  \label{eq:Bayes}
\eeq
which here
relates the \emph{posterior} $\pdf(x|y,I)$ to the \emph{likelihood} $\pdf(y|x,I)$
given the \emph{prior} $\pdf(x|I)$ and the \emph{evidence} $\pdf(y|I)$.
These relations will enable us to derive the posterior for $\Delta_k$
in terms of assumed priors.

Another implication of the product rule follows when $x$ and $y$ are
mutually independent, which means that knowing $y$ doesn't affect the
probability of $x$, so that $\pdf(x|y,I) = \pdf(x|I)$. Then
Eq.~\eqref{eq:product} tells us that
\beq
  \pdf(x,y|I) \longrightarrow \pdf(x|I)\,\pdf(y|I)
  \;.
  \label{eq:independent}
\eeq

In the following we sometimes omit the explicit $I$, but the specification of
prior information should always be assumed.

\subsection{CH synopsis and EFT correspondence}

We consider multiple priors that reflect the expectation that all coefficients
in the expansion of an observable in powers of $Q$ are of roughly the same size---or, more
precisely, they have a distribution with a characteristic size.
The fundamental assumption made by Cacciari and Houdeau in
Ref.~\cite{Cacciari:2011ze}
is that all coefficients of $\alpha_s$ in the pQCD series are
roughly the same size, which is implemented by treating them as
random variables having a shared
distribution with upper bound $\cbar$.
This assumption is motivated by empirical evidence from the behavior of such series.
But it may not be correct for EFT expansions, where the form (\ref{eq:obsexp})
is expected to result in coefficients which are $\mathcal{O}(1)$, not arbitrarily large.

\begin{table}[tb]
  \caption{Prior pdfs.}
  \label{tab:priors}
  \begin{tabular}{c|cc}
  set  &   $\pdf(c_n| \cbar)$   &   $\pdf(\cbar)$  \\
  \hline
  A &
  $\frac{\ts 1}{\ts 2\cbar}\,\theta(\cbar-|c_n|)$
    & $\frac{\ts 1}{\ts \ln \cbarmax/\cbarmin}
   \frac{\ts 1}{\ts\mathstrut\cbar}
    \theta ( \cbar - \cbarmin) \theta( \cbarmax - \cbar)$
   \\[8pt]
  B &
  $\frac{\ts 1}{\ts 2\cbar}\,\theta(\cbar-|c_n|)$
  &  $\frac{\ts 1}{\ts\sqrt{2\pi}\cbar\sigma} e^{-(\log\cbar)^2/2\sigma^2}$
   \\[8pt]
  C &
  $\frac{\ts 1}{\ts\sqrt{2\pi}\cbar} e^{-c_n^2/2\cbar^2}$
    & $\frac{\ts 1}{\ts \ln \cbarmax/\cbarmin}
   \frac{\ts 1}{\ts\mathstrut\cbar}
    \theta ( \cbar - \cbarmin) \theta( \cbarmax - \cbar)$
  \\[5pt]
  \hline
  \end{tabular}
\end{table}

To proceed we need to translate such a fundamental assumption into concrete
expressions for priors.
Cacciari and Houdeau do this through three supplementary assumptions
(which they call ``hypotheses'')~\cite{Cacciari:2011ze}, as follows.
\begin{itemize}
  \item
  The prior probability densities for coefficients at different orders are
   independent in the sense of \eqref{eq:independent}. I.e., given an upper bound $\cbar$, the joint prior density
  for coefficients factorizes:
	\beq
		\pdf(c_0, \ldots, c_n | \cbar) = \prod_{i=0}^n \pdf (c_i | \cbar)
		\;.
		\label{eq:factorize}
	\eeq
  CH then also assume that $\pdf(c_i|\cbar)$ is the same pdf for each $i$.	
  Thus the value of $\cbar$ is the most
  knowledge obtainable from the known coefficients when predicting
  possible values of unknown ones.  In this way, we have isolated communication from the data about the sum of all omitted higher-order terms into one variable $\cbar$.

  \item
  Next we need a specific prior probability distribution for
  $\pdf(c_i|\cbar)$. In the interest of understanding the prior-dependence of our analysis, we test alternative implementations of naturalness in the priors.
The extent to which the posterior pdf for $\Delta_k$ is stable under different, but reasonable, choices of prior indicates the extent to which data on $\{c_n:n=0, \ldots, k\}$ is sufficiently informative to dominate the analysis.

 When we know there is an upper bound to the coefficients, an application of maximum entropy dictates that the least-informative distribution is uniform for $|c_i| < \cbar$.
 Such uniformity is additionally appealing because it can lead to simple, analytic results.
 This uniform prior is the initial choice of Ref.~\cite{Cacciari:2011ze}. We employ it in priors we denote as ``Set A" and ``Set B" (see Table~\ref{tab:priors}), the difference being the prior pdf assumed for
 $\cbar$ in the two cases (see below).

The priors of ``Set C" in Table~\ref{tab:priors} then correspond to the ensemble
  naturalness assumption of Ref.~\cite{Schindler:2008fh}. This Gaussian prior follows
  from the maximum-entropy principle assuming knowledge of testable information on the mean and standard deviation of the $c_n$'s~\cite{Sivia:06,Gull:98}
  \beq
    \label{eq:ensnat_conditions}
     \left\langle \sum_{n=0}^{k} c_n^2 \right\rangle = (k+1)\cbar^2, \quad \left\langle c_n \right\rangle = 0
     \;.
  \eeq

We will see below that analyses with Sets A and B are insensitive to details of the distribution of $\{c_0, \ldots, c_k\}$: the only feature of the distribution that matters is the value of the largest of these $k+1$ lower-order coefficients. On the other hand, results under Set C priors are affected by the specific distribution of these coefficients, as well as by the largest value.

  \item
  Finally, the application of Bayes' theorem requires a prior for
 $\cbar$: $\pdf(\cbar)$.
Uniformity of $\ln\cbar$ is the only way to ensure unbiased expectations regarding the scale of $\cbar$~\cite{Jeffreys:1939}.
 This log-uniform prior for $\cbar$ was chosen in Ref.~\cite{Cacciari:2011ze}, 
 and so Set A of Table~\ref{tab:priors} is their choice of prior. 
 We also employ the log-uniform prior for $\cbar$ in Set C, there following Schindler 
 and Phillips in Ref.~\cite{Schindler:2008fh}.  Such a prior cannot be normalized 
 for $\cbar$ in $(-\infty,\infty)$ and is therefore termed an ``improper prior". 
 Limiting the range of $\cbar$ through the use of $\theta$ functions 
 permits an examination of the otherwise ill-defined limiting behavior.
CH chose $\cbarmin = \epsilon$ and $\cbarmax = 1/\epsilon$
  and take the limit $\epsilon\rightarrow 0$ at the end of the calculations.  Thus, 
  the $\theta$ functions and associated $\ln\cbarmax/\cbarmin$ factor serve to regulate the 
  distribution so that the pdf is
  always normalized. Taking the limit $\epsilon \rightarrow 0$ expresses complete 
  ignorance of the scale of $\cbar$, although we will also consider finite 
  ranges for the marginalization over $\cbar$, thereby rendering the priors more 
  informative.
  	
  In Refs.~\cite{Bagnaschi:2014wea,Bagnaschi:2014eva}, a more informative
  $\cbar$ prior is considered based on the fact that the first
  coefficient $c_0$ can be scaled out. The authors argue that in this case it is no longer necessary to allow for an arbitrarily large value for the other coefficients.
  In consequence they assume $\cbar$'s prior is
  a log-normal distribution about zero. We take this as Set~B of
  Table~\ref{tab:priors}. Note that scaling the observable so that the first coefficient is order unity is also what we do for the EFT expansion, see Eq.~(\ref{eq:obsexp}).
\end{itemize}

In the case of prior information on the naturalness of coefficients that is different than that discussed here, maximum entropy can be used to derive how the different information should be reflected in alternative priors~\cite{Gull:98}.  Such direct conversion from information on the interpretation of naturalness to prior pdfs facilitates rigorous derivation of the consequences of the concepts in question through the use of formal reasoning and the language of probability. We now show how this transpires by deriving the pdf for $\Delta_k$, initially refraining from specifying anything about the priors on $\{c_0, \ldots, c_k\}$.

\subsection{Posteriors and DOB intervals for $\Delta_k$}
Given the three assumptions described above and the prior sets of Table~\ref{tab:priors},
we can systematically derive the posterior for $\Delta_k$ by repeated
application of the sum and product rules and their logical Bayesian consequences.
At each step, we introduce a specific concept being built into the analysis.
For this general derivation we assume that
the coefficients start from $k=0$ and are all significant and non-zero (later we will modify the general results to treat the case of NN scattering in chiral EFT, where the orders are nominally $\Q^0$, $\Q^2$, $\Q^3$, \ldots with $\Q^1$ absent).

\begin{widetext}
    \begin{enumerate}

      \item \textbf{Formula for $\pr\mathbf{(\Delta_k|c_0,...,c_k)}$:}
       We seek $\pdf(\Delta_k| c_0,\ldots,c_k)$, which is the probability density
    for the dimensionless residual, $\Delta_k$, given the known values of the first
    $k+1$ coefficients.  Because the true value of $\Delta_k$ depends only (explicitly)
    on the unknown coefficients
    $c_{n>k}$,
    we insert them into the equation by
    integrating over all their possible values using \eqref{eq:marginalize}
    and \eqref{eq:product},
    %
    %\begin{widetext}
    \begin{align}
      \pdf(\Delta_k|c_0,\ldots,c_k) &=
        \int\!
        \pdf(\Delta_k|c_{k+1},c_{k+2},\ldots)
             \, \pdf(c_{k+1},c_{k+2},\ldots | c_0,\ldots,c_k) \,
                   dc_{k+1} \, dc_{k+2}\, \cdots
      \nonumber \\
      &= \int \left[
      \delta\left(\Delta_k - \sum_{n=k+1}^\infty c_n \Q^n \right)
      \right]
     \, \pdf(c_{k+1},c_{k+2},\ldots | c_0,\ldots,c_k) \,
      dc_{k+1} \, dc_{k+2}\, \cdots \;,
      \label{eq:deltaeq}
    \end{align}
    %\end{widetext}
    %
    where we have used
    \beq
      \pdf(\Delta_k|c_0,\ldots,c_k,c_{k+1},c_{k+2},\ldots)
      =
      \pdf(\Delta_k|c_{k+1},c_{k+2},\ldots)
       =
      \delta\left(\Delta_k - \sum_{n=k+1}^\infty c_n \Q^n \right)  \;.
    \eeq
    The latter is a direct consequence of Eq.~\eqref{eq:Deltak_def}.
    Equation~\eqref{eq:deltaeq} states that, to get a specified $\Delta_k$ given a set of known coefficients,
    we need to sum up all the different
    combinations of $c_n$'s with $n > k$ that give us this $\Delta_k$, weighting each combination by
    its probability given the known values of coefficients
    $c_n$ for $n\leq k$. Note that all of these integrals over $c_n$ are from $-\infty$ to $+\infty$
    in general, but in particular cases there may be constraints (e.g., a cross section
    is positive, so the leading coefficient will be positive).

    The probability density \eqref{eq:deltaeq} is
    correctly normalized since the normalization integral over $\Delta_k$ can be performed using the delta function,
    leaving the normalization integral for $\pdf(c_{k+1},c_{k+2},\ldots | c_0,\ldots,c_k)$,
    which is unity.

    \item \textbf{Independent priors:}
    Our priors are based on the assumption that
    $\cbar$ is the only information that gets transmitted to the
    distribution of $c_{n}$ for $n>k$.
    Thus, at this stage we introduce $\cbar$ as an intermediary in the pdf in
    the integrand of Eq.~\eqref{eq:deltaeq}
    via another marginalization integral, and apply this assumption:
    %
    %\begin{widetext}
    \begin{align}
      \pdf(c_{k+1},c_{k+2},\ldots | c_0,\ldots,c_k)
      &= \int_0^\infty \!
       \pdf(c_{k+1},c_{k+2},\ldots | \cbar, c_0,\ldots,c_k) \, \pdf(\cbar | c_0,\ldots,c_k)\, d\cbar
     \nonumber \\
      &= \int_0^\infty \!
       \pdf(c_{k+1},c_{k+2},\ldots | \cbar) \, \pdf(\cbar | c_0,\ldots,c_k)\, d\cbar
     \nonumber \\
       &= \int_0^\infty
       \Bigl[\prod_{n=k+1}^{\infty} \pdf(c_n | \cbar)\Bigr]
          \pdf(\cbar | c_0,\ldots,c_k)\, d\cbar
        \;.
        \label{eq:cbarfirst}
    \end{align}
    %\end{widetext}
    %
    In the final equality we have used the assumption that the
    $c_i$ distributions are independent, causing the joint densities for the $c_n$'s with $n>k$ to become the product
    of independent densities $\pdf(c_n|\cbar)$.
    We will see that the imposition of Set A or Set C priors for $\cbar$ makes the limits on this integral finite.

    \item
    \textbf{Leading-term approximation:} We next assume 
    $\Delta_k \approx c_{k+1}\Q^{k+1} \equiv \Deltakone$
    in Eq.~\eqref{eq:deltaeq} and return later to relax this assumption.
    By examining the effect of this assumption on DOB intervals, we will show that this approximation is quite appropriate for small values of $\Q$.
    This is the simplest way to
    exploit the delta function, which then depends only on $c_{k+1}$.
    After substituting Eq.~\eqref{eq:cbarfirst} into \eqref{eq:deltaeq}, the
    $c_{k+2},\ldots$ integrals are just normalization integrals (equal to one),
    leaving integrations over $c_{k+1}$ and $\cbar$:
    %
    %\begin{widetext}
    \begin{align}
     \pdf(\Deltakone|c_0,\ldots,c_k) &= \int_{-\infty}^\infty \int_0^\infty \bigl[
       \delta(\Deltakone -  c_{k+1} \Q^{k+1})
      \bigr]
      \pdf(c_{k+1}|\cbar) \, \pdf(\cbar | c_0,\ldots,c_k)\, d\cbar\, dc_{k+1}
      %\label{eq:intermediate}
      \nonumber
      \\
      &= \frac{1}{\Q^{k+1}} \int_0^\infty \!
        \pdf(c_{k+1} = \Deltakone/\Q^{k+1} | \cbar) \,
        \pdf(\cbar | c_0,\ldots,c_k)\, d\cbar
        \;. \label{eq:ckplusone}
    \end{align}
    %\end{widetext}
    %

    \item \textbf{Expanding the composite prior:}
    The first pdf in the integrand of
    \eqref{eq:ckplusone} may be directly evaluated for a given choice of prior,
    but the second cannot. It can, however, be identified as being constructed 
    from the priors defined in Table~\ref{tab:priors}, via
    application of Bayes' theorem:
    \begin{align}
       \feps(\cbar | c_0,\ldots,c_k) &=
       \frac {\feps(c_0,\ldots,c_k | \cbar)\, \feps(\cbar)} {\feps(c_0,\ldots,c_k)}
       \nonumber \\  &=
           \frac {\feps(c_0,\ldots,c_k | \cbar)\, \feps(\cbar)}
           {\int_0^\infty \feps(c_0,\ldots,c_k | \cbar')\, \feps(\cbar')\, d\cbar'}
       \nonumber \\ &=
           \frac {\ds\Bigl[\prod_{n=0}^{k} \pr(c_n | \cbar)\Bigr] \feps(\cbar)}
           {\ds\int_0^\infty \Bigl[\prod_{n=0}^{k} \pr(c_n | \cbar')\Bigr] \feps(\cbar')\, d\cbar'}
       \;.
       \label{eq:BayesEps}
    \end{align}
    In the second line
    we have introduced another marginalization over
    $\cbar'$ in the denominator, while in the third line we apply the independence
    assumption of Eq.~\eqref{eq:factorize}.
    Combining \eqref{eq:BayesEps} and \eqref{eq:ckplusone} gives:
    \beq
     \pdf(\Deltakone|c_0,\ldots,c_k) =
       \frac{ \ds\int_0^\infty \!
        \pdf(c_{k+1} = \Deltakone/\Q^{k+1} | \cbar) \,
        \ds\Bigl[\prod_{n=0}^{k} \pr(c_n | \cbar)\Bigr] \feps(\cbar)
        \, d\cbar}
        {\ds \Q^{k+1} \int_0^\infty \Bigl[\prod_{n=0}^{k} \pr(c_n | \cbar')\Bigr] \feps(\cbar')\, d\cbar'}
       \;.
       \label{eq:BayesEps2}
    \eeq
   Now we can apply one of the sets of assumptions in Table~\ref{tab:priors},
    which give us specific forms
    to evaluate each of the pdfs in Eq.~\eqref{eq:BayesEps2}.
    Note that if some of the $c_i$'s for $i < k$ are identically zero, there are
    correspondingly fewer terms in the products of $\pr(c_n | \cbar)$ in
    Eq.~\eqref{eq:BayesEps2}.

    \item \textbf{Prior Set $\Aone$:}
    Prior Set~A has been developed under the assumption
    that identifying a maximum value $\cbar$ is a valid
    concept.
    We here start with a finite $\cbar$ range between $\cbarmin$ and $\cbarmax$ for which Eq.~\eqref{eq:BayesEps2} can be evaluated analytically.
    If $\cbarmin = \epsilon$ and $\cbarmax = 1/\epsilon$
    and we take the limit as $\epsilon \rightarrow 0$, we designate this as Set $\Aeps$. Meanwhile, the superscript $^{(1)}$ is introduced to denote
    the use of the first-term approximation.

    For this prior choice, $\Aone$, the denominator of Eq.~\eqref{eq:BayesEps2} is directly evaluated
    as there are only integrals over theta functions:
    %
    %\begin{widetext}
    \begin{align}
       \int_0^\infty \Bigl[\prod_{n=0}^{k} \pdf(c_n|\cbar') \Bigr] \,
       \feps(\cbar')\, d\cbar'
      &= \int_0^\infty
          \Bigl[\prod_{n=0}^{k} \frac{1}{2\cbar'}\theta(\cbar'-|c_n|) \Bigr]
            \frac{1}{\ln\cbarmax/\cbarmin} \,\frac{1}{\cbar'} \, \theta(\cbar'-\cbarmin)
             \theta(\cbarmax - \cbar')
          \, d\cbar'
        \nonumber \\
      &= \frac{1}{2^{k+1}}\frac{1}{\ln\cbarmax/\cbarmin}
         \int_{\max(\cbark,\cbarmin)}^{\cbarmax} \frac{1}{\cbar'{}^{k+2}}
         \, d\cbar'
         \;,
    \end{align}
    where we have followed CH~\cite{Cacciari:2011ze} and introduced the variable  $\cbar_{(j)}$ to denote the maximum of the first $j+1$ coefficients:
    \beq
       \cbar_{(j)} \equiv \max(|c_0|,\cdots,|c_j|)
       \;.
       \label{eq:cbarkDef}
    \eeq
    %\end{widetext}
    %
    The integration over $\cbar$ in the numerator of Eq.~\eqref{eq:BayesEps2} is similar, but contains the
    extra pdf $\pdf(c_{k+1} = \Deltakone/\Q^{k+1} | \cbar)$ in the
    integrand.  Using the theta functions to once again define the integration bounds, the numerator simplifies to
    \beq
     \frac{1}{2^{k+2}} \frac{1}{\ln \cbarmax/\cbarmin} \theta\left( \cbarmax - \cbarkpone\right) \int_{\max(\cbarkpone, \cbarmin)}^{\cbarmax} \frac{1}{\cbar^{k+3}}  \, d \cbar \;.
    \eeq
    We can see now how regulating the integrals with $\cbarmax$ and $\cbarmin$ will allow
    terms such as $\ln\cbarmax/\cbarmin$ (and most factors of 2)
    to cancel between the numerator and denominator, after which we may choose to
    take $\epsilon$ to zero in $\Aepsone$ without consequence.

More generally, we assume that the integration range for $\cbar$ is wide enough that $\cbarmin < \cbark < \cbarmax$. The posterior then evaluates to:
%\begin{widetext}
  \beq
   \pr(\Deltakone|\Data) =
   \frac{1}{\Q^{k+1}} \frac{1}{2}  \left(\frac{k+1}{k+2}\right)
   \frac{\theta \left( \cbarmax - \cbarkpone \right)}{ \cbark^{-(k+1)} - \cbarmax^{-(k+1)}}
   \Scale[0.95]{
  \begin{cases}
    \cbark^{-(k+2)} - \cbarmax^{-(k+2)}
    & \mbox{if }\left| \Deltakone \right| \leq \cbark \Q^{k+1}
    \vspace{2mm}
      \\
    \left(\frac{\ts\Q^{k+1}}{\ts\bigl|\Deltakone\bigr|}\right)^{k+2} - \cbarmax^{-(k+2)}
    & \mbox{if }\left| \Deltakone \right| > \cbark \Q^{k+1}
   \end{cases}
   }
  \;.
    \eeq
    %\end{widetext}
   If some of the coefficients are zero  (e.g., the series
  starts at $\Q^l$ with $l>0$, or one or more
  intermediate coefficients are zero) we can revise these formulas trivially:
   the only change is that there are fewer theta functions in the
  integrals.  Taking $n_c$ to be the number of non-zero constants, we implement
  this generalization by replacing
  $k$ by $n_c-1$ everywhere except for powers of $\Q$, which remain
  $k+1$.  Thus, the modified posterior for Set $\Aone$ is
	%
	%\begin{widetext}
  \beq
  \hspace{7mm}
  \label{eq:setAposterior}
   \pr(\Deltakone|\Data) =
   \frac{1}{\Q^{k+1}} \frac{1}{2}  \left(\frac{n_c}{n_c+1}\right)
   \frac{\theta \left( \cbarmax - \cbarkpone \right)}{\cbark^{-n_c} - \cbarmax^{-n_c}}
   \Scale[0.95]{
  \begin{cases}
    \cbark^{-(n_c+1)} - \cbarmax^{-(n_c + 1)}
    & \mbox{if }\left| \Deltakone \right| \leq \cbark \Q^{k+1}
    \vspace{2mm}
      \\
    \left(\frac{\ts\Q^{k+1}}{\ts\bigl|\Deltakone\bigr|}\right)^{n_c+1} - \cbarmax^{-(n_c+1)}
    & \mbox{if }\left| \Deltakone \right| > \cbark \Q^{k+1}
   \end{cases}
   }
   \;,
    \eeq
    %\end{widetext}
    which simplifies to the corresponding equation of Ref.~\cite{Cacciari:2011ze} in the limiting case of $\Aepsone$
    \beq
          \pdf(\Deltakone | c_0, \ldots, c_k) =
      \left(\frac{n_c}{n_c+1} \right)
      \frac{1}{2\cbark\Q^{k+1}}
      \begin{cases}
        1 & \mbox{if }|\Deltakone| \leq \cbark\Q^{k+1} \\
        \left(\frac{\ts\cbark\Q^{k+1}}{\ts\bigl|\Deltakone\bigr|}\right)^{n_c+1}
        & \mbox{if }|\Deltakone| > \cbark\Q^{k+1}
      \end{cases}
    \;.
    \eeq
    Note that this simple generalization is possible due to the identical treatment of the priors for each coefficient.

    \item \textbf{Prior Sets $\Bone$ and $\Cone$:}
        Neither Set~B nor Set~C priors allow for analytic integrals over $\cbar$, so the discussion here will necessarily be less extensive than that for Set~A.
        In the first-term approximation the posterior for $\Delta_k$ can be reduced to a one-dimensional integral, whose evaluation must be left to numerical integration.
        Inserting Set~C priors into Eq.~\eqref{eq:BayesEps2} results in
        \beq
        \label{eq:setCposterior}
          \pr(\Deltakone|\Data) =
          \frac{
          \frac{\ts 1}{\ts\Q^{k+1}} \frac{\ts 1}{\ts\sqrt{2\pi}} 
          \int\limits_{\cbarmin}^{\cbarmax} d \cbar \, 
          \exp{\left[-\frac{\left(\Deltakone\right)^2}
              {2 \left(\Q^{k+1}\right)^2\cbar^2} \right]} 
              \left(\frac{\ts 1}{\ts\strut\cbar}\right)^{n_c + 2} 
              \left[ \prod\limits_n e^{-c_n^2/2\cbar^2} \right]
          }{
          \int\limits_{\cbarmin}^{\cbarmax} d \cbar'\,\left(\frac{\ts 1}{\ts\cbar'}\right)^{n_c + 1}\left[ \prod\limits_n e^{-c_n^2/2\cbar'^2} \right] 
          }
          \;,
        \eeq
        where products are assumed to run over all $n_c$ coefficients with defined prior distributions.

    \item \textbf{DOB intervals: }
        In the  Bayesian framework, the posterior $\pdf(\Delta_k | c_0, \ldots, c_k)$
        contains the complete
        information we claim to have about the dimensionless residual $\Delta_k$.
        In some applications we need to use the entire posterior because it is
        very structured (e.g., multi-modal or simply non-gaussian), but here
        we can capture most of the information with the choice of a small number
        of degree-of-belief (DOB) intervals.%
        \footnote{These are also called ``credibility" or ``credible" intervals.}

        In particular, the DOB for a particular interval in $\Delta_k$ is
        found simply by integrating $\pdf(\Delta_k | c_0, \ldots, c_k)$
        over this interval.  We could also start with a given DOB, e.g., the standard frequentist (gaussian) 68\% or 95\%, and determine the smallest interval that integrates
        to that number.  Or we could specify some other criterion for deciding the interval,
        such as that it is symmetric about the mode.
        In fact, the use of any of the priors
        in Table~\ref{tab:priors} results in a smallest p\%-DOB interval
        for $\Delta_k$ that is symmetric about the mode; following Ref.~\cite{Cacciari:2011ze} we
        denote the corresponding dimensionless limits by $\pm d_k^{(p)}$.
        Thus the implicit definition of this interval is
        \beq
            p\% = \int_{-d_k^{(p)}}^{d_k^{(p)}} \pdf(\Delta_k | c_0, \ldots, c_k)
            \, d\Delta_k
            \;.
            \label{eq:integralford}
        \eeq

        In the limiting case of prior Set~$\Aepsone$, this integral can be evaluated
        explicitly~\cite{Cacciari:2011ze}:
        \beq
            d_k^{(p)} = \cbark \, \Q^{k+1}
            \times
            \begin{cases}
            \frac{\ts n_c+1}{\ts n_c}\, p\%
            & \mbox{if } p\ \leq \frac{\ts n_c}{\ts n_c+1}  \\
            \Bigl[ \frac{\ts 1}{\ts (n_c+1)(1-p\%)} \Bigr]^{1/n_c}
              & \mbox{if } p\ > \frac{\ts n_c}{\ts n_c+1}
            \end{cases}
            \label{eq:dcoeffs}
        \;,
        \eeq
        where $n_c$ is again the number of non-zero known coefficients.  
        Thus, with these priors, the interval of 
        width $\cbark \Q^{k+1}$ about the EFT prediction at order $k$ is a 
        $n_c/(n_c + 1)*100\%$ DOB interval, cf.\  Ref.~\cite{Cacciari:2011ze}. 
        Such a theory error bar has often been
        assigned in previous EFT calculations, and---as we shall discuss further 
        in Section~\ref{sec:EKM}---corresponds to the prescription formalized in 
        Refs.~\cite{Epelbaum:2014efa,Epelbaum:2014sza}. It is important---e.g., 
        in the context of error propagation---to keep in mind that this prior 
        leads to a distribution of probability for the truncation error that is 
        not Gaussian.

        For the more general form of prior Set~$\Aone$, an analytic formula
        for $d_k^{(p)}$ can still be found.
        The explicit form of the integral depends on the $p\%$ value of interest, 
        because of the change in structure for 
        $\bigl| \Deltakone \bigr| > \cbark \Q^{k+1}$.
        Thus, we first calculate this transition value $\ppercentt$ by integrating 
        the maximal probability within the first region in which 
        $|c_{k+1}| \leq \cbark$ to obtain
        \beq
           \ppercentt %= \int\limits_{0}^{\cbark \Q^{k+1}} \frac{1}{\Q^{k+1}} \left( \frac{n_c}{n_c+1}\right) \left[ \frac{\frac{1}{\cbark^{n_c+1}} - \frac{1}{\cbarmax^{n_c+1}}   }{\frac{1}{\cbark^{n_c}} - \frac{1}{\cbarmax^{n_c}}} \right] \dif \Deltakone
           =  \left[ \frac{\frac{1}{\strut\cbark^{n_c+1}} - \frac{1}{\strut\cbarmax^{n_c+1}}
           }{
           \frac{1}{\strut\cbark^{n_c}} - \frac{1}{\strut\cbarmax^{n_c}}} \right]  \left(\frac{n_c}{n_c+1}\right) \cbark
          \;.
          \label{eq:pregion1}
        \eeq
        Now suppose we are interested in $p\%$ intervals for $p < p_t$. 
        Equation~\eqref{eq:pregion1} implies that the interval bounded by variation $\pm \cbark \Q^{k+1}$ is a \mbox{$\ppercentt$-DOB interval}.  Generally,
         the DOB interval for Set~$\Aone$ is bounded by
        \beq
            \label{eq:aonedkpregion1}
            \dkp =\left[ \frac{\frac{1}{\strut\cbark^{n_c+1}} - \frac{1}{\strut\cbarmax^{n_c+1}}
           }{
           \frac{1}{\strut\cbark^{n_c}} - \frac{1}{\strut\cbarmax^{n_c}}} \right]^{-1} \frac{n_c+1}{n_c}\, p\% \, \Q^{k+1} \qquad \mbox{if } p\% \leq \ppercentt
        \;.
        \eeq
        When one is interested in larger $p\%$ values, it may be beneficial to take advantage of the normalization of the pdf to conduct an integration in only one region by integrating the second case of Eq.~\eqref{eq:setAposterior} on the interval $\left[\dkp,\infty\right]$.
        Because $\cbarkpone = c_{k+1}$ in this region, the theta function truncates this integration at $\Deltakone = \cbarmax \Q^{k+1}$.
        The resulting implicit expression for $\dkp$ if $p\% > \ppercentt$ is thus
        \begin{eqnarray}
        \label{eq:aonedkpregion2}
        (1-p\%)=\frac{1}{\Q^{k+1}} \left(\frac{n_c}{n_c+1}\right) \frac{1}{ \frac{1}{\strut\cbark^{n_c}} -  \frac{1}{\strut\cbarmax^{n_c}}} \left[ \frac{\left(\dkp - \cbarmax \Q^{k+1}\right)}{\cbarmax^{n_c+1}}  +
        \frac{\left(\Q^{k+1}\right)^{n_c+1}}{n_c} \left(\frac{1}{\bigl(\dkp\bigr)^{n_c}} - \frac{1}{(\cbarmax \Q^{k+1})^{n_c}} \right) \right].\nonumber\\
        \end{eqnarray}

       For Set~$\Bone$, Set $\Cone$, or, indeed, for any of the sets if we do not make the first-term approximation,
       the DOB interval $\dkp$ can be found numerically from Eq.~\eqref{eq:integralford}
       by integrating $\pdf(\Delta_k | c_0, \ldots, c_k)$
       (e.g., from Eq.~\eqref{eq:setCposterior}) from zero until 
       the integral equals $p/2$.  
        We stress again the resulting DOB intervals are
        \emph{not} standard 
        deviations and make no statement about the shape of the normalized 
        function which integrates to 0.68 between the bounds $\pm d_k^{(68)}$.

        % One advantage of working with coefficients scaled by a value on the order of the coefficients themselves $\left\{ c_n \right\}$ is that the the region of support of the $\cbar$ integral is reduced and the infinite integral becomes less computationally expensive.
        % It may be additionally advantageous to present the integration after a change of variables which isolates the $\Q$ dependence to the integration bounds.
        % Such modifications may be expressed in terms of a root-finding procedure as
        % \beq
        %     \label{eq:dkpCone}
        %     0 = \left[ \int_{-\frac{\dkp}{X_0\Q^{k+1}}}^{\frac{\dkp} {X_0\Q^{k+1}}}
        %      d c_{k+1} \, \frac{ \frac{1}{\sqrt{2\pi}} \int\limits_{\cbarmin}^{\cbarmax} d \cbar \, \exp{\left[-\frac{c_{k+1}^2}{2\cbar^2} \right]} \left(\frac{\ts 1}{\ts\cbar}\right)^{k+3} \left[ \prod\limits_{n = 0}^k e^{-c_n^2/2\cbar^2} \right]
        %     }{
        %     \int\limits_{\cbarmin}^{\cbarmax} d \cbar'\,\left(\frac{\ts 1}{\ts\cbar'}\right)^{k+2}\left[ \prod\limits_{n = 0}^k  e^{-c_n^2/2\cbar'^2} \right]
        %     } - p\% \right]
        %     \;,
        % \eeq
        %Note that once the $\cbar$ integration is completed, the pdf is no longer Gaussian.
        
    \item \textbf{Relaxation of first-term approximation:}
    To relax the assumption that the first omitted term dominates, we introduce the generalized notation
    \beq
        \Delta_k \approx
        \Deltakone + \sum_{m=k+2}^{\kmax} c_m \Q^m \equiv \Delta_k^{(\kmax-k)} = \Delta_k^{(h)} \;,
        \label{eq:tildeDeltak}
    \eeq
    where $\kmax$ is the highest-order coefficient kept in the sum of omitted
    terms.
    Returning to step 3 above,
    we continue to use the $\delta$ function to eliminate the integral over
    $c_{k+1}$, with the result that $\Deltakone$ is replaced by $\Delta_k^{(\kmax-k)}$ in
    the subsequent expression and the integrations over $c_{m}$ for $m > k+1$
    up to $m = \kmax$ remain. The generalization of Eq.~\eqref{eq:BayesEps2} is then
    \beq
     \pdf(\Delta_k^{(h)}|c_0,\ldots,c_k) =
       \frac{ \ds \int_{-\infty}^\infty \!\cdots\!\int_{-\infty}^\infty \int_0^\infty\!
        \pdf(c_{k+1} = \Delta_k^{(h)}(\Q)/\Q^{k+1} | \cbar) \,
        \ds\Bigl[\prod_{n=0}^{k} \pr(c_n | \cbar)\Bigr] \feps(\cbar)
        \, d\cbar \, \prod_{m=k+2}^{\kmax} \pr(c_m | \cbar)\,dc_m }
        {\ds \Q^{k+1} \int_0^\infty \Bigl[\prod_{n=0}^{k} \pr(c_n | \cbar')\Bigr] \feps(\cbar')\, d\cbar'}
       \;,
       \label{eq:BayesEpsFull}
    \eeq
    where there are $\kmax - k - 1$ integrals from $-\infty$ to $\infty$ in the numerator, in addition to the integral over the (positive) $\cbar$.
    Note that if the first omitted term really does dominate,
    %then $\Delta_k^{(1)} \rightarrow \Delta_k$ 
    %for a small value of $h$ 
    then the integrals over higher
    $c_m$'s are trivial normalization integrals, restoring the result of the first-omitted-term approximation.

    \item \textbf{Summary:}
    We have derived a general result for $\pdf(\Delta_k|c_0,\ldots,c_k)$
    in Eq.~\eqref{eq:BayesEpsFull},
    which is valid for any of the sets in Table~\ref{tab:priors}.
    In most cases this expression must be evaluated numerically, for example by Monte Carlo
    integration.  By assuming the first omitted term dominates, we obtain
    the much less involved integration in Eq.~\eqref{eq:BayesEps2}.  Evaluating the
    application of this approximation to Set~$\Aone$ yields the analytic result 
    in Eq.~\eqref{eq:setAposterior} while for Sets $\Bone$ and $\Cone$ integrals are left to be evaluated numerically---see, e.g.,
    Eq.~\eqref{eq:setCposterior}.
    Finally, DOB intervals can be derived from these posteriors 
    analytically for $\Aone$ (Eqs.~\eqref{eq:dcoeffs}, \eqref{eq:aonedkpregion1}, 
    and \eqref{eq:aonedkpregion2}) 
    and numerically for the others from Eq.~\eqref{eq:integralford}.
\end{enumerate}

\end{widetext}

\subsection{Representative examples}

\begin{figure*}[tbh!]
  \includegraphics[width=0.32\textwidth]{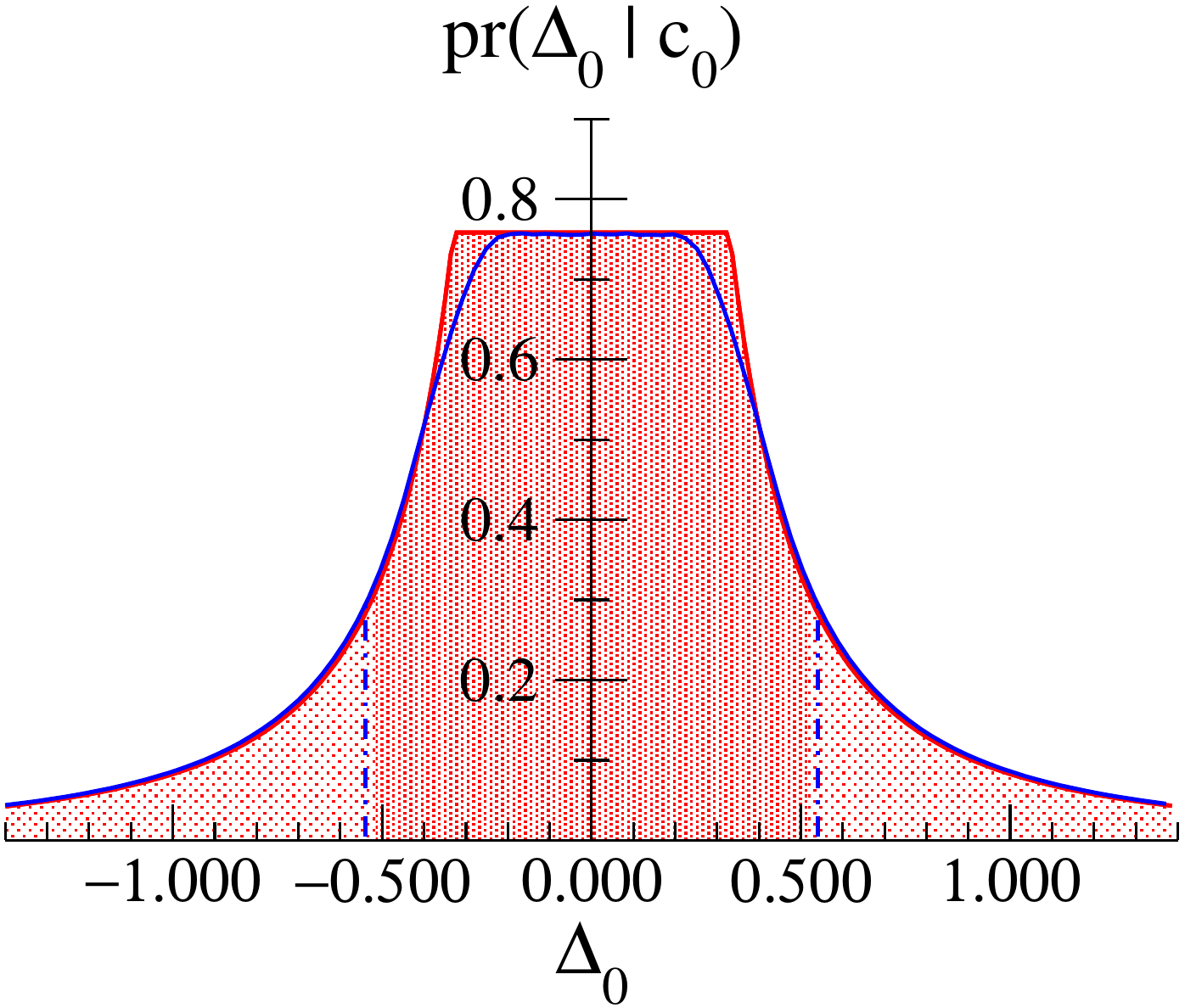}~~%
  \includegraphics[width=0.32\textwidth]{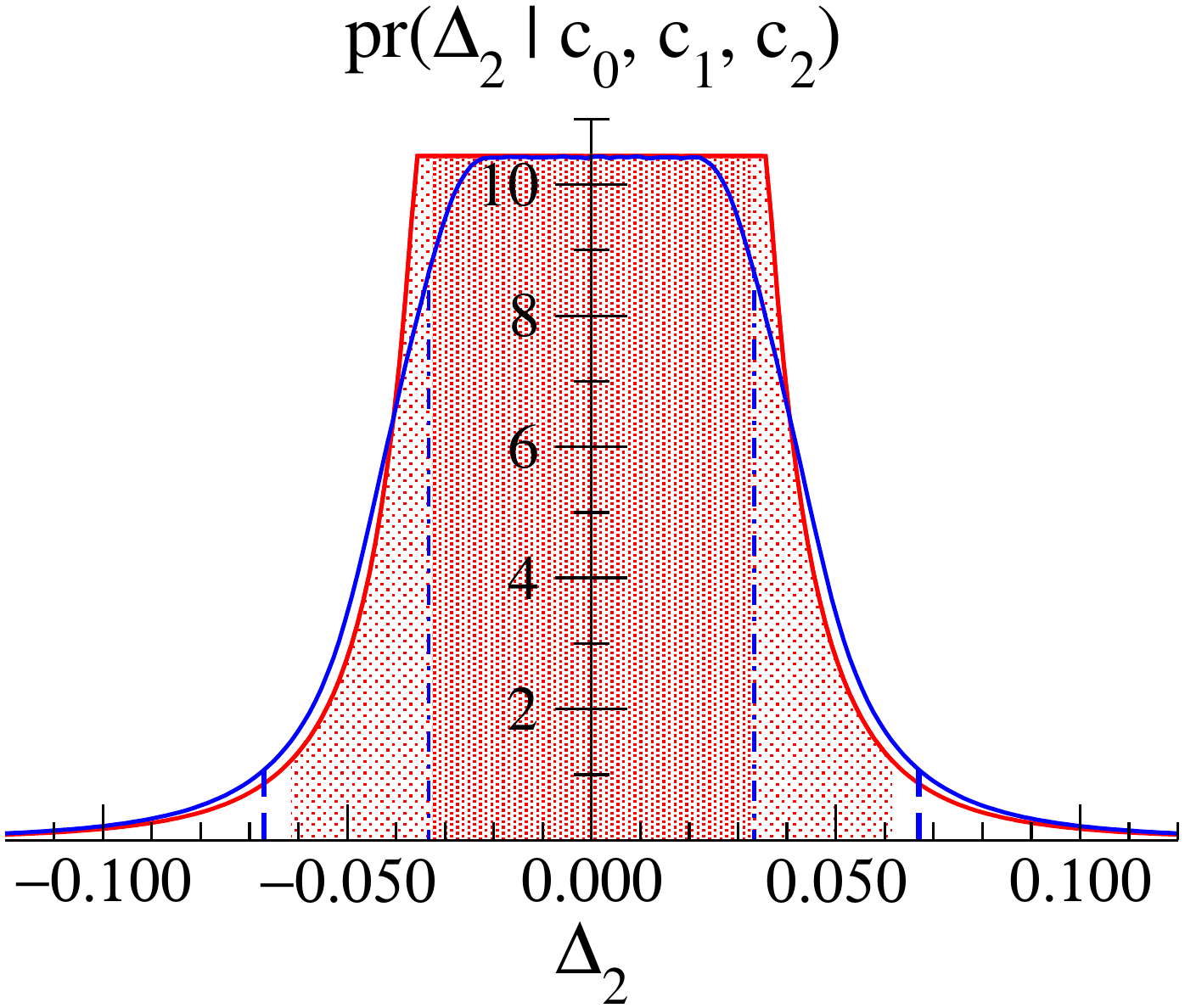}
  \includegraphics[width=0.32\textwidth]{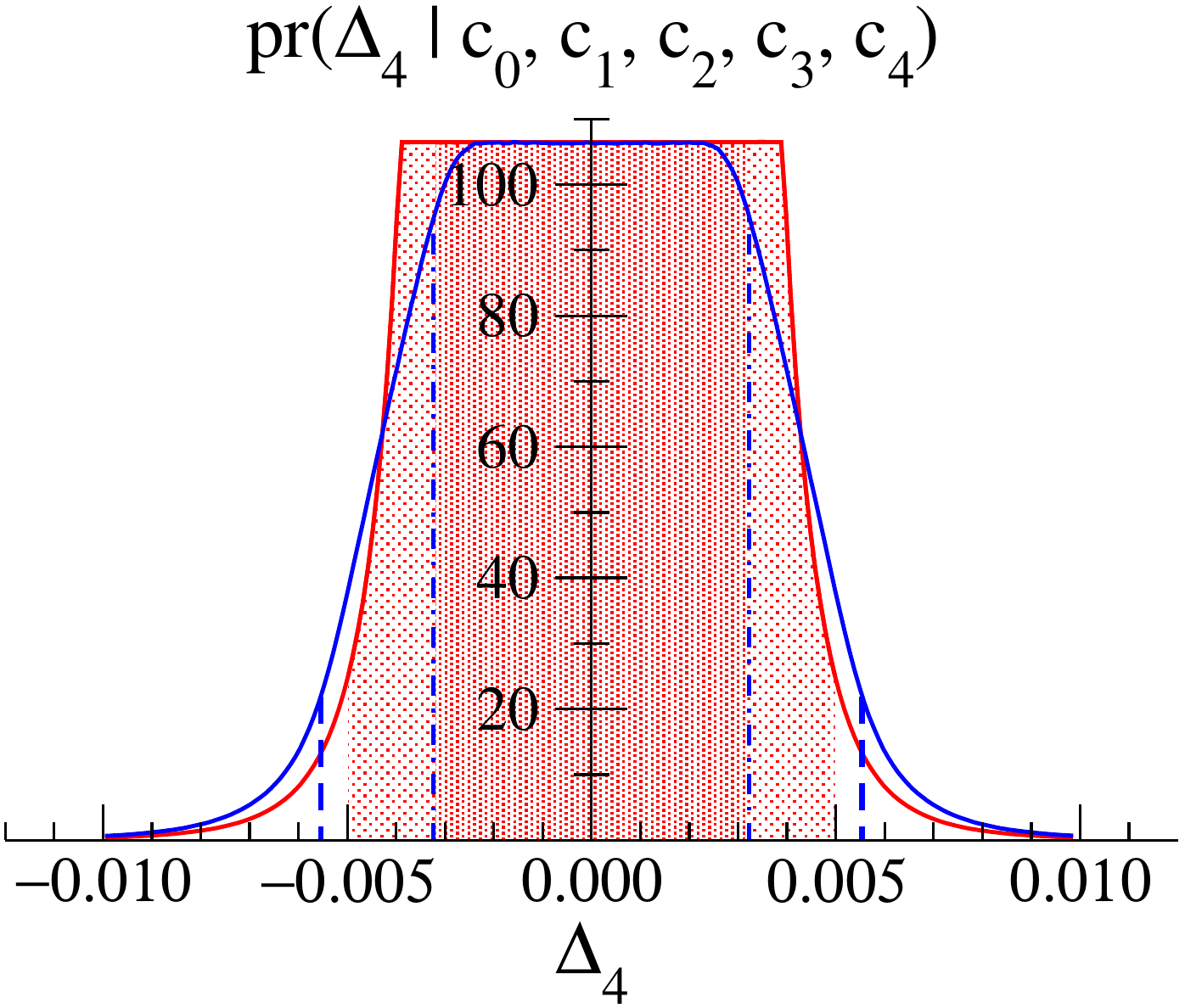}
 \caption{(color online) Posteriors for $\Deltakone$ under Set $\Aepsone$ priors of Table~\ref{tab:priors}
   for $k=0$, 2, and 4 with $\Q=0.33$.  In all cases, $c_n = 1$ was assumed.
   The solid red curve is the analytic result from Eq.~\eqref{eq:setAposterior}, with the shaded regions
   marking the 68\% and 95\% DOB intervals.  The solid blue curve is the posterior $\Delta_k$ for
   $\Aeps$ once higher-order terms are included, with dot-dashed and dashed lines marking
   the corresponding 68\% and 95\% DOB intervals.}
 \label{fig:setAposterior}
\end{figure*}

\begin{figure*}[tbh!]
 \includegraphics[width=0.32\textwidth]{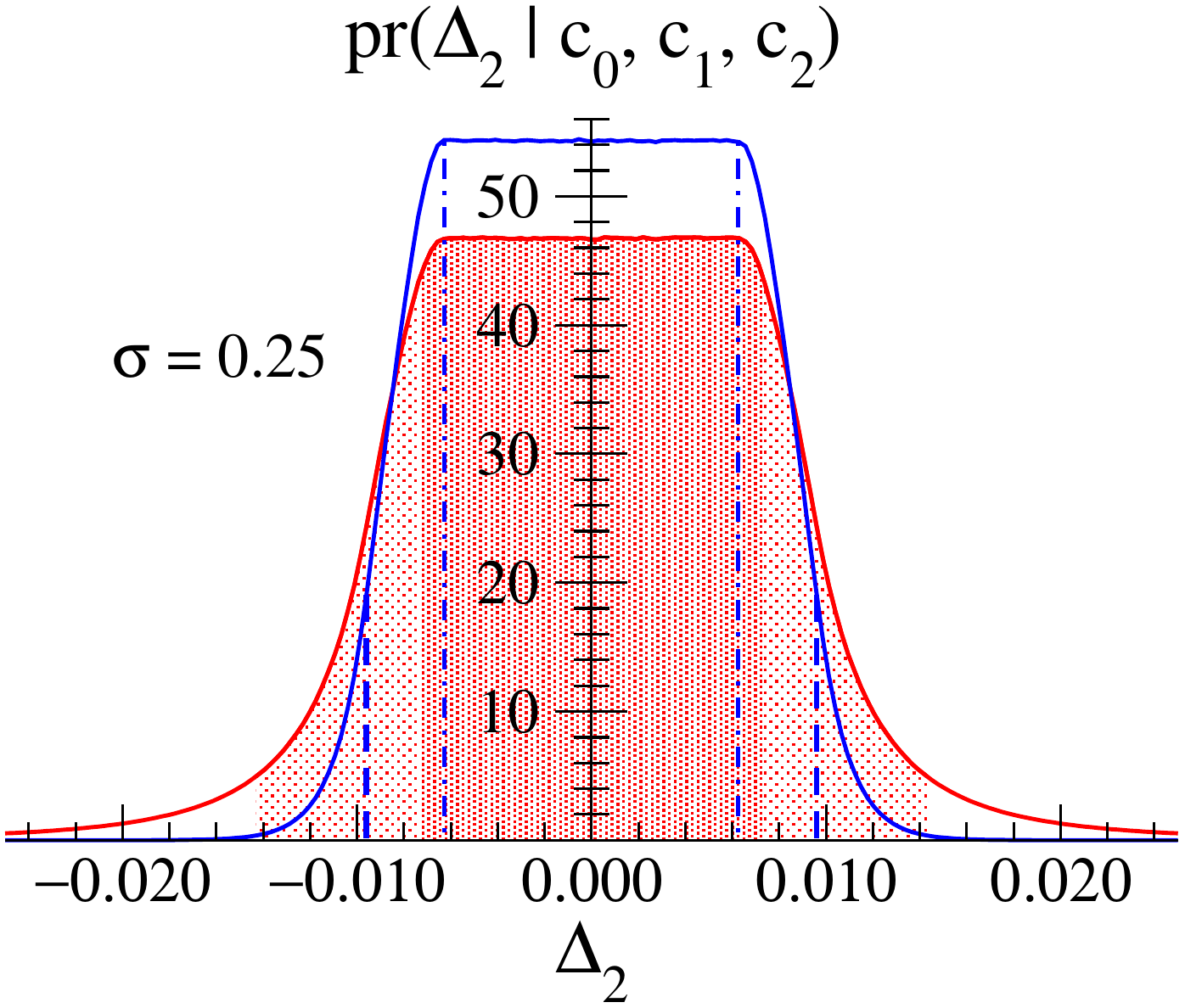}~%
 \includegraphics[width=0.32\textwidth]{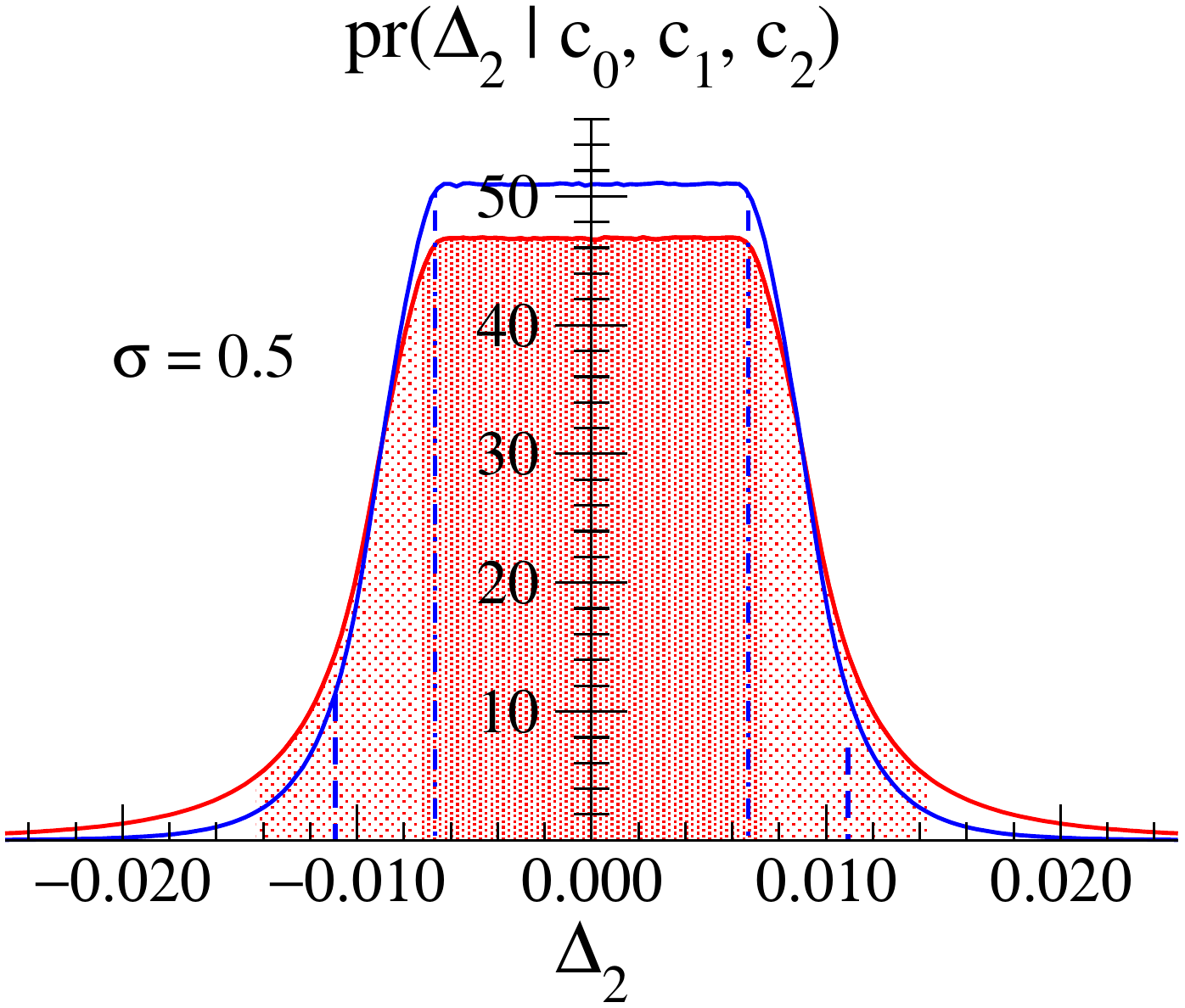}~%
 \includegraphics[width=0.32\textwidth]{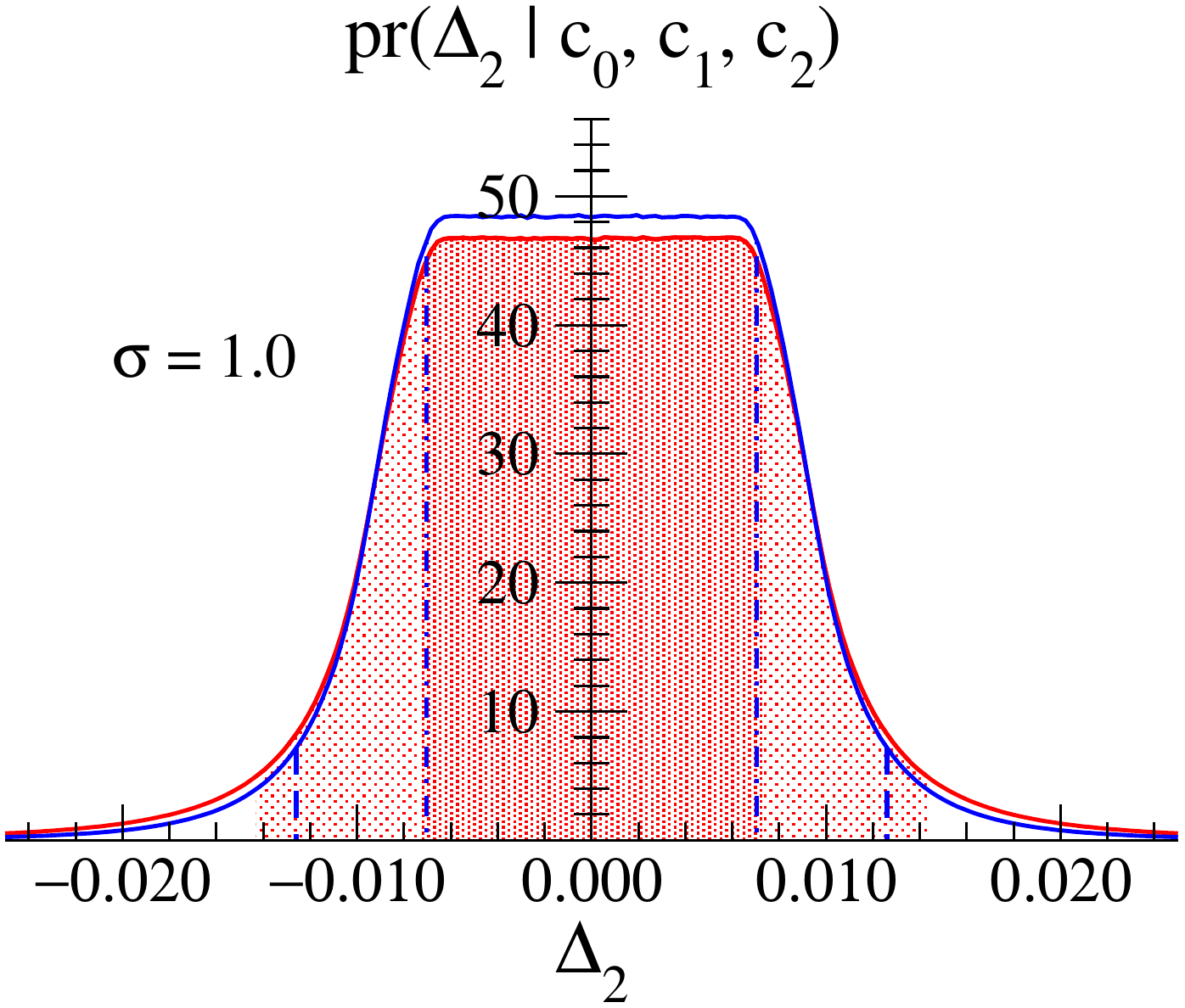}~%
  \caption{(color online) Comparison of $\Delta_2$ for prior set \Aeps\ (red solid line) and
  B (blue solid line) for $\sigma=0.25$, $0.5$ and $1.00$ respectively. In each case $c_0,c_1,c_2$ are all set to
 unity with an expansion parameter $\Q = 0.2$. The DOB intervals are indicated as
 in Fig.~\ref{fig:setAposterior}.}
 \label{fig:compareABposteriors0p20}
\end{figure*}

\begin{figure*}[tbh!]
 \includegraphics[width=0.32\textwidth]{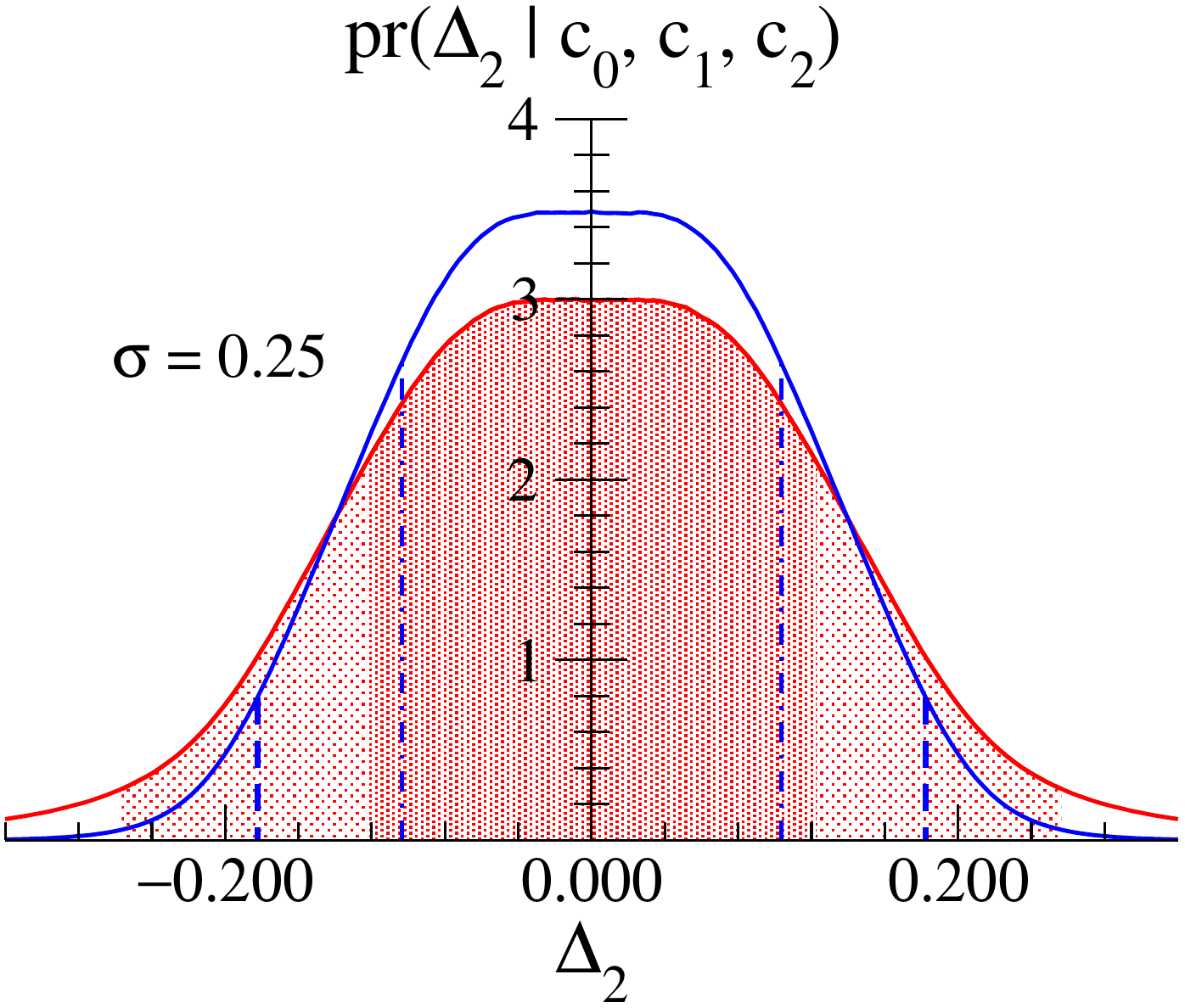}~%
 \includegraphics[width=0.32\textwidth]{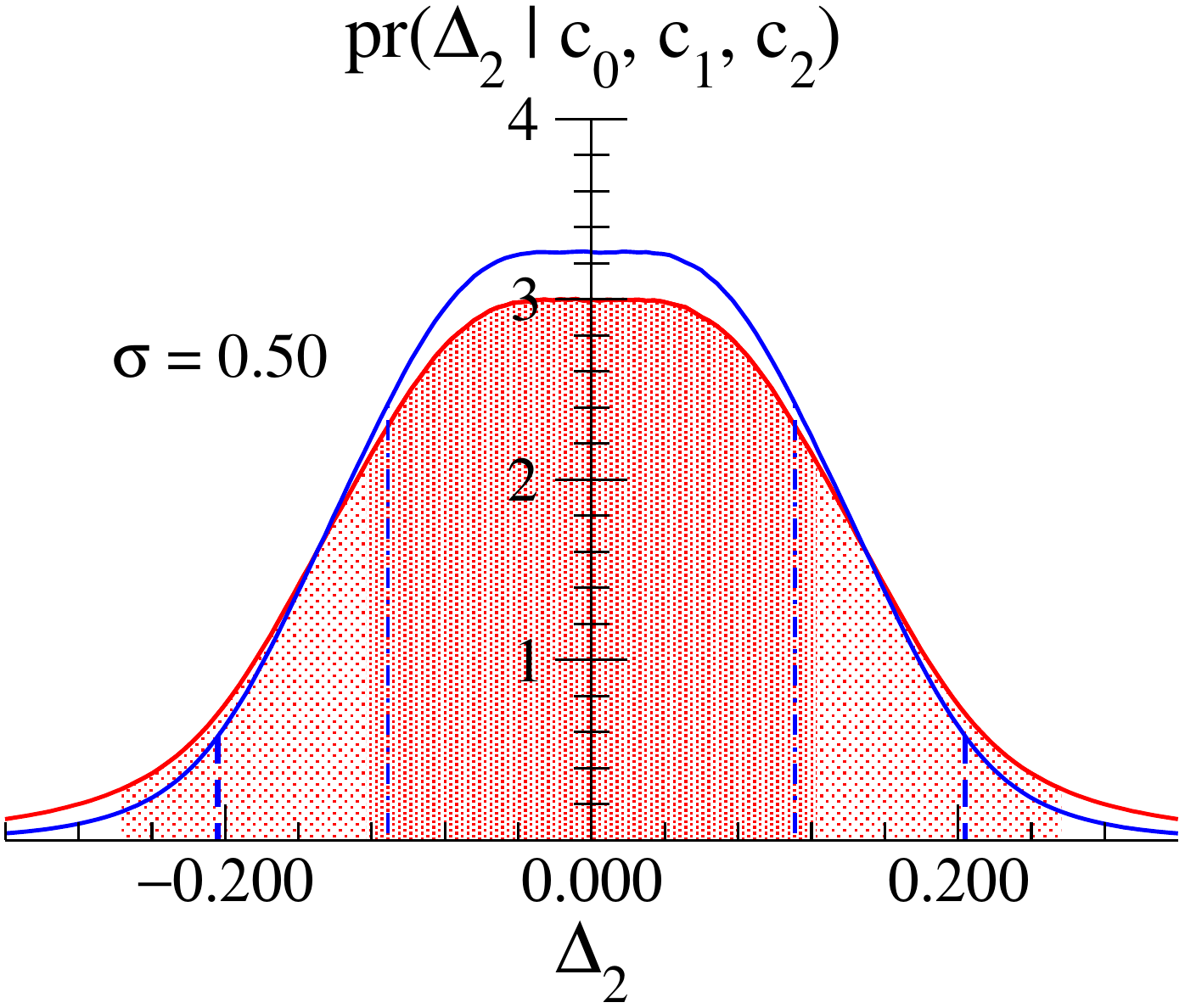}~%
 \includegraphics[width=0.32\textwidth]{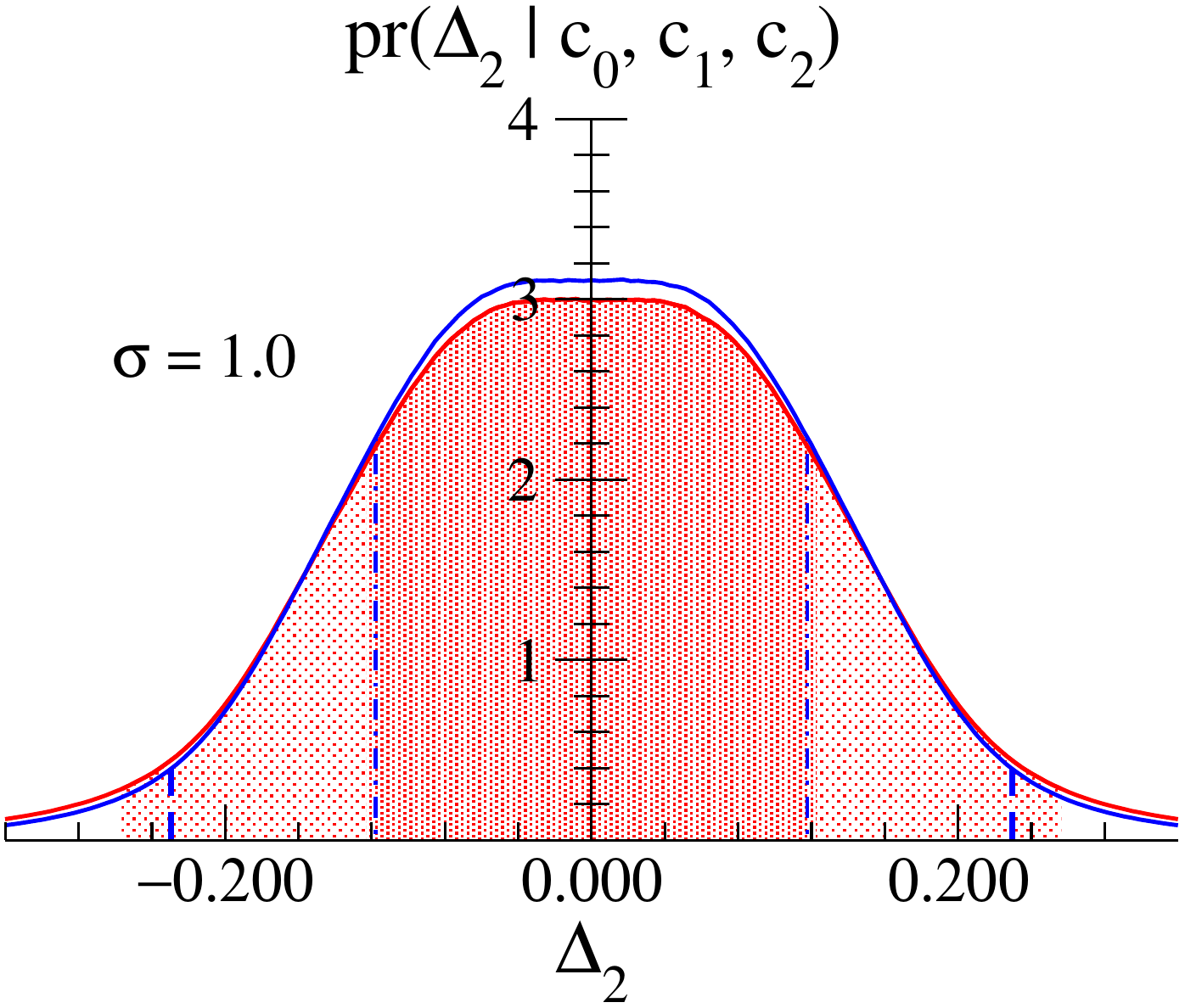}~%
  \caption{(color online) Comparison of $\Delta_2$ for prior set \Aeps\ (red solid line) and
  B (blue solid line) for three values of $\sigma$ and $c_0,c_1,c_2$ all set to
 unity with an expansion parameter $\Q = 0.5$.
 The DOB intervals are indicated as in Fig.~\ref{fig:setAposterior}.}
 \label{fig:compareABposteriors0p50}
\end{figure*}

\begin{figure*}[tbh!]
 \includegraphics[width=0.32\textwidth]{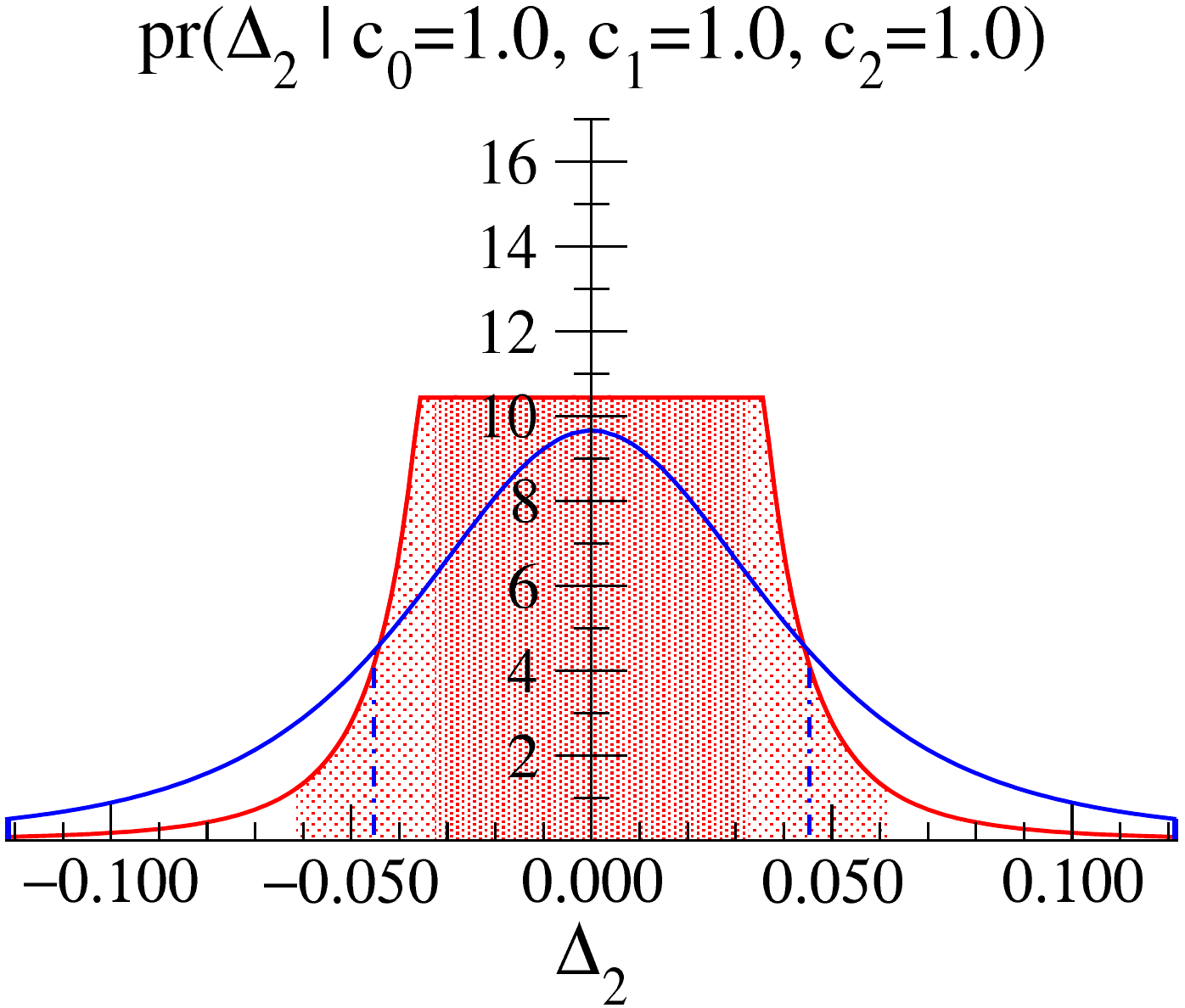}~%
 \includegraphics[width=0.32\textwidth]{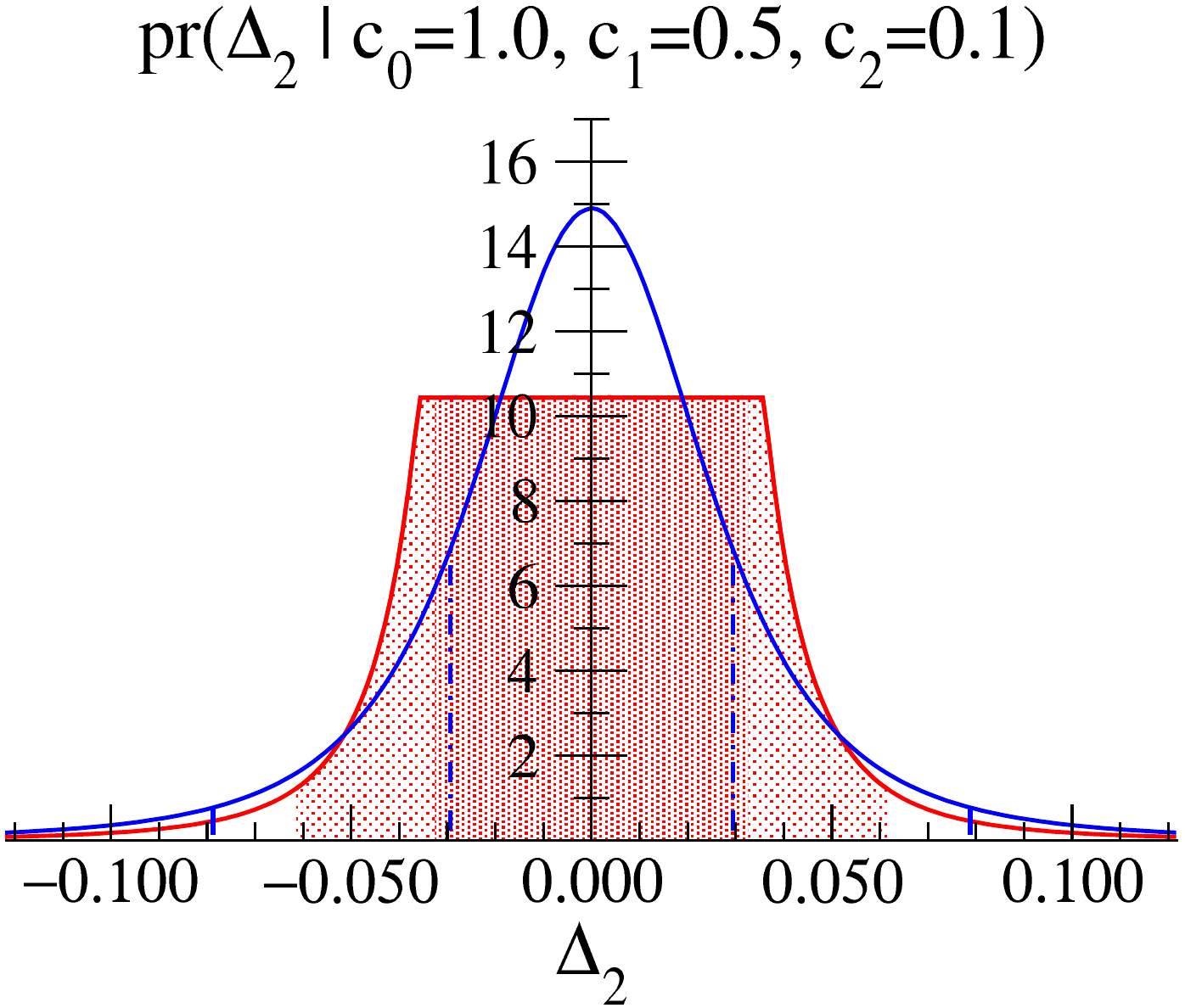}~%
 \includegraphics[width=0.32\textwidth]{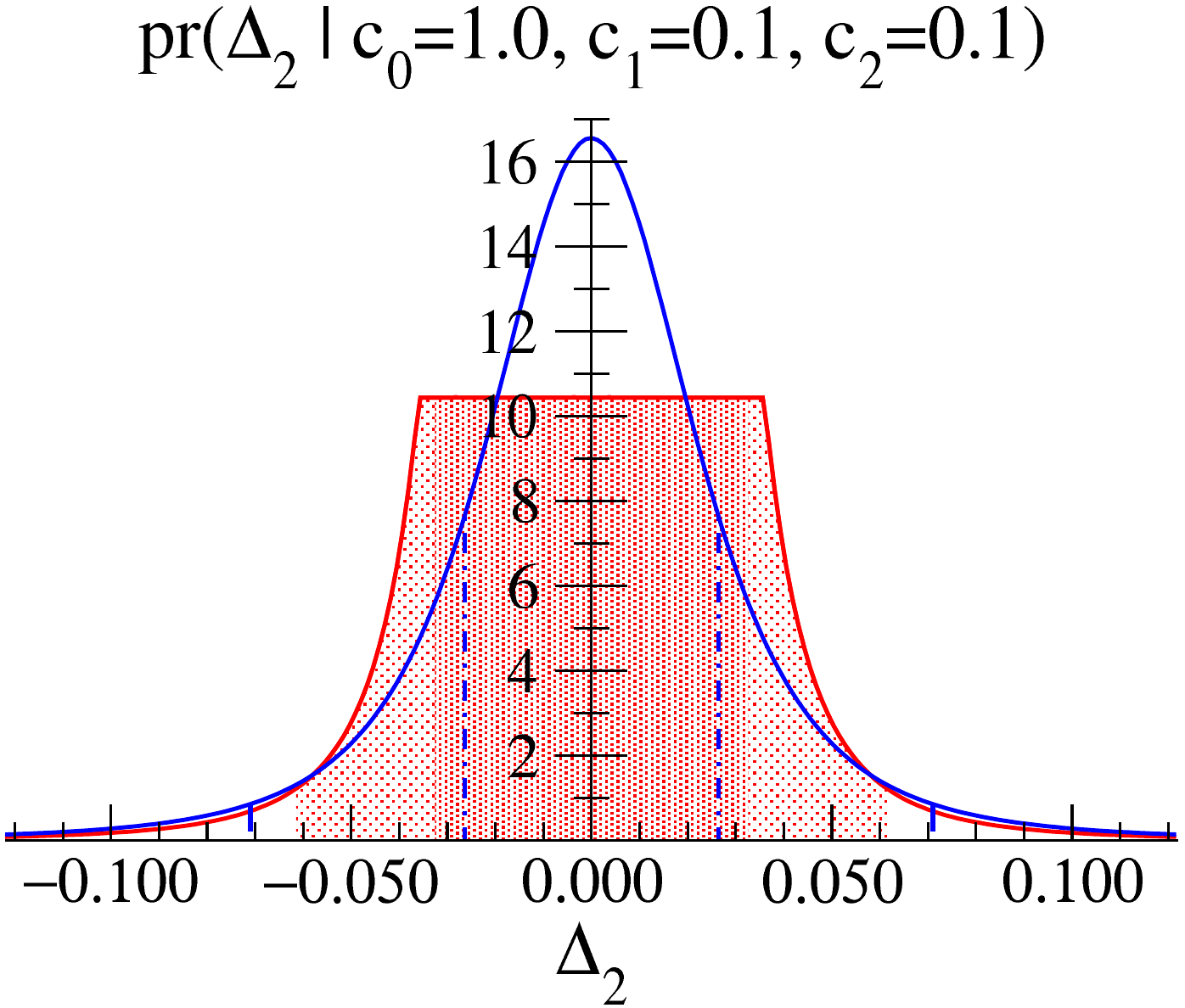}~%
  \caption{(color online) Comparison of $\Delta_2$ for prior set \Aeps$^{(1)}$\ (red solid line)
  and \Ceps$^{(1)}$\ (blue solid line)---note both are in the leading-omitted-term
  approximation---for three sets of $c_n$ values with an expansion parameter $\Q = 0.33$.
  These sets are (left-to-right) $\{c_n\}(a)\equiv \{1.0,1.0,1.0\}$, $\{c_n\}(b) \equiv\{1.0,0.5,0.1\}$,
  and $\{c_n\}(c) \equiv \{1.0,0.1,0.1\}$.
 The DOB intervals are indicated as in Fig.~\ref{fig:setAposterior}.}
 \label{fig:compareACposteriors0p33}
\end{figure*}

Before applying the Bayesian framework developed by CH, and extended above,
to the specific problem of NN scattering, we make some
general observations on the form of the posteriors for $\Delta_k$
and the systematics of the
68\% and 95\% DOB intervals for various prior sets from
Table~\ref{tab:priors}.

We start with the set \Aeps, defined by $\cbarmin = \epsilon$,
$\cbarmax = 1/\epsilon$, with $\epsilon \rightarrow 0$
(in practice all results here in which $\epsilon$ is invoked use $\epsilon = 0.001$).
The posterior distribution for $\Deltakone$ in Eq.~\eqref{eq:setAposterior},
which assumes the first omitted term dominates the error,
has a flat central plateau with power-suppressed tails.
This is illustrated by the red curves in Fig.~\ref{fig:setAposterior} for
$k=0$, $k=2$, and $k=4$, for $\Q=0.33$.  The heavy and light shaded regions show
the 68\% and 95\% DOB intervals, respectively.
From Eq.~\eqref{eq:dcoeffs}, the width of the posterior is given by
$\cbark \Q^{k+1}$ times a number of order unity, so the dominant effect is that
the width decreases by a factor of $\Q$ with each increase of $k$ by one.
In fact, only the maximum value of the $c_n$'s for a given $k$, $\cbark$, matters under this choice of prior; the distribution of those $c_n$'s is irrelevant.
The overall size of all DOB intervals then scales linearly with $\cbark$, so here $c_0$, \ldots, $c_k$ have all been set to one for simplicity.
For Set A priors the generalization to other cases is trivial.

To relax the first-term approximation we include the first four omitted terms in our computation of $\Delta_k$.
The result is then converged numerically in all cases shown, so in practice Set $Y^{(4)}$ ($Y=A$, $B$, or $C$) priors lead to the same results as
when arbitrarily many higher-order terms included in the truncation-error calculation. In consequence, we do not include superscripts below
when reporting results with terms beyond the first omitted one included in the computation of $\Delta_k$. Such calculations show
that the central plateau in the posterior becomes rounded (blue curve
in Fig.~\ref{fig:setAposterior}).
The corresponding effect on the DOB intervals depends on the value
of $k$; for $k\geq 2$ there is no significant
effect on the 68\% DOB intervals while the 95\% intervals are increased slightly.

If we use the more informative log-normal prior for $\cbar$ from set B, the tails are
more quickly suppressed than for set \Aeps, to a degree that depends in detail on the values of
$\sigma$ and $\Q$~\cite{Bagnaschi:2014eva,Bagnaschi:2014wea}.
Representative examples for $\Delta_2$
are shown in Figs.~\ref{fig:compareABposteriors0p20}
and \ref{fig:compareABposteriors0p50} for $\Q=0.2$ and $\Q=0.5$, respectively,
with $\sigma=0.25$, $0.5$, and $1.0$ respectively.
We see that the 68\% DOB and 95\% DOB intervals are smaller that those for \Aeps,
with the difference increasing with smaller $\sigma$.
The further extension of the tail for \Aeps\ is not surprising
as we have allowed for the possibility of $\cbar$ having a large range.
As $\sigma$ gets larger, the posteriors for each $\Q$ value
approach the \Aeps\ result;  once
$\sigma \geq 1.0$ there is very little difference between $\Aeps$ and Set B for $k\geq 2$. $k=0$ and $k=1$ are more sensitive
to $\sigma$.

\begin{table*}[tbh]
 \setlength{\tabcolsep}{9pt}
 \caption{Resulting 68\% and 95\% DOB intervals for $\Delta_k$ using Set $\Aone$ with different choices for the minimum and maximum of the $\cbar$ prior. In all cases $\cbark = 1$.}
 \label{tab:setA_approx2}
  \begin{tabular}{c|c|c|c|c|c|c|c|c}
  &  min/max & $\Q$ & $k=0$ & $k=1$ & $k=2$ & $k=3$ & $k=4$  \\
  \hline
     & 0.001/1000 &  &
   0.31 & 0.041 & 0.0073 & 0.00136 & 0.00026 \\
     & 0.25/4.0 & 0.20  & 0.22 & 0.039 & 0.0072 & 0.00136 & 0.00026 \\
     & 0.50/2.0 &     &  0.18 & 0.035 & 0.0068 & 0.00132 & 0.00026 \\
  \cline{2-8}
     & 0.001/1000 &   &  0.51 & 0.111 & 0.033 & 0.0101 & 0.0032 \\
   68\%  & 0.25/4.0 & 0.33    & 0.36 & 0.106 & 0.032 & 0.0101 & 0.0032 \\
     & 0.50/2.0 &   &   0.30 & 0.095 & 0.030 & 0.0098 & 0.0031 \\
  \cline{2-8}
     & 0.001/1000 & & 0.78 & 0.26 & 0.113 & 0.053 & 0.026 \\
     & 0.25/4.0 & 0.50    &   0.55 & 0.243 & 0.112 & 0.053 & 0.025 \\
     & 0.50/2.0 &    &   0.45 & 0.22 & 0.106 & 0.051 & 0.025 \\
  \hline\hline
    & 0.001/1000 &   &  1.96 & 0.103 & 0.0137 & 0.0023 & 0.00041 \\
    & 0.25/4.0 & 0.20    &  0.47 & 0.077 & 0.0129 & 0.0022 & 0.00041 \\
    & 0.50/2.0 &    &  0.29 & 0.056 & 0.011 & 0.0020 & 0.00039 \\
  \cline{2-8}
    & 0.001/1000 &   &  3.2 & 0.28 & 0.0614 & 0.0168 & 0.0050 \\
   95\%  & 0.25/4.0 & 0.33    &   0.77 & 0.21 & 0.058 & 0.0166 & 0.0050 \\
    & 0.50/2.0 &    &   0.48 & 0.152 & 0.048 & 0.0150 & 0.0047 \\
  \cline{2-8}
    & 0.001/1000 &  &  4.91 & 0.645 & 0.21 & 0.0884 & 0.040 \\
    & 0.25/4.0 & 0.50    &  1.16 & 0.48 & 0.201 & 0.087 & 0.040 \\
    & 0.50/2.0 &    &  0.73 & 0.35 & 0.166 & 0.079 & 0.038 \\
  \hline
 \end{tabular}
\end{table*}

We might expect that the Set B results with $\sigma > 1.0$ will be even closer to those from
Set A if we impose a range of $\cbar$ values in Set A that
reflects naturalness expectations.
Results of varying the range over which $\cbar$ is marginalized are shown in Table~\ref{tab:setA_approx2}, where
we compare DOB intervals for $\Delta_k$ with $k$ from 0 to 4.
For $k\geq 3$, the change in the range of $\cbar$ has no noticeable effect on either the 68\% or 95\%
DOB.
For $k=2$, effects are 5--10\% on the 68\% interval if a narrow range ($\cbar$ from 0.5 to 2.0) is employed.  Effects on the 95\%
interval can be up to 20\% if this narrow range is employed at $k=2$.

\begin{table*}[tbh]
 \setlength{\tabcolsep}{9pt}
 \caption{DOB intervals for Sets \Aeps\ and \Ceps,
 with and without the leading omitted term
 approximation, for $k=2$, with three sets of $c_n$ values.
 These sets are
 $\{c_n\}(a)\equiv \{1.0,1.0,1.0\}$, $\{c_n\}(b) \equiv\{1.0,0.5,0.1\}$,
  and $\{c_n\}(c) \equiv \{1.0,0.1,0.1\}$. }
 \label{tab:results_setC_approx1}
 \begin{tabular}{c|c|c|c|c|c|c}
   & $\Q$ & $\Aepsone/\Aeps$ & $\Cepsone/\Ceps\ \{c_n\}(a)$ & $\Cepsone/\Ceps\ \{c_n\}(b)$
       & $\Cepsone/\Ceps\ \{c_n\}(c) $  \\
  \hline
    & 0.20 & 0.0073/0.0073  & 0.0095/0.0097  & 0.0062/0.0063  & 0.0056/0.0057  \\
   68\%   & 0.33 & 0.033/0.033  & 0.043/0.045  & 0.028/0.029  & 0.025/0.026  \\
     & 0.50 & 0.113/0.123  & 0.149/0.171  & 0.096/0.111  & 0.087/0.100  \\
 \hline\hline
      & 0.20 & 0.0137/0.0137   & 0.025/0.026   & 0.017/0.017   & 0.015/0.015  \\
   95\%   & 0.33 & 0.061/0.066  & 0.114/0.121   & 0.074/0.079  & 0.067/0.071 \\
     & 0.50 & 0.21/0.25  & 0.40/0.46  & 0.26/0.30   & 0.23/0.27 \\
   \hline
 \end{tabular}
\end{table*}

The Set~C priors are qualitatively different from Set~A because they
correspond to an ensemble naturalness assumption for $\pdf(c_i|\cbar)$,
which means that the distribution of $c_n$'s for a given $k$---and not just their
maximum---affects the result.
This is illustrated by the results in Table~\ref{tab:results_setC_approx1},
in which DOB intervals  for $\Delta_2$ with prior choices  \Aeps\ and \Ceps\
are compared.
Because $k=2$, the coefficients $c_0$, $c_1$, and $c_2$
are all influential; we consider three representative choices for their values.
The systematics going from $c_n$ sets $a$ to $b$ to $c$ show that having more $c_n$'s near one leads to larger DOB intervals.
Taking $c_n(b)$ to give generic results for a roughly even distribution of coefficients we find that  the Set C vs.\ Set A comparison is
close: only a 10-15\% increase for the 68\% DOB and a roughly 20\% increase for
the 95\% DOB. (The Set A intervals are wider for all but the case in which all three known coefficients are 1.0.) This reflects the stronger central peaking of the Set C pdf under a reasonable distribution of the first three coefficients, as depicted (in the first-omitted-term approximation)
in Fig.~\ref{fig:compareACposteriors0p33}.
Such differences in DOB intervals under different prior choices will be amplified if $k=0$ or $k=1$.

Table~\ref{tab:results_setC_approx1} also assesses the approximation of keeping only the leading omitted
term in $\Delta_2$. Once $\Q=0.5$ we see appreciable differences
between Set $\Cone$ and Set C results that include multiple higher-order terms,
but even then it is only about a 15\% effect on the error bar.

%%%%%%%%%%%%%%%%%%%%%%%%%%%%%%%%%%%%%%%%%%%%%%%%%%%%%%%%%%%%%%

\section{Comparison to recent chiral EFT results for np scattering} \label{sec:EKM}

\subsection{EKM's truncation-error estimates}

Chiral perturbation theory ($\chiPT$) encodes the consequences of QCD at
momenta of order the pion mass~\cite{Weinberg:1978kz,Gasser:1983yg,Jenkins:1990jv,Bernard:2006gx}. It can be used to compute the
interaction of single nucleons and pions for momenta well below the
chiral-symmetry-breaking scale, $\LamchiSB$. $\chi$PT yields a purely perturbative expansion in powers of
$(p,m_\pi)/\LamchiSB$ for low-energy pion-pion and pion-nucleon scattering. But, nuclei are bound
states, and will not be generated from such an expansion.

In the early 1990s Weinberg pointed out that the infrared enhancement associated
with multi-nucleon intermediate states meant that the $\chi$PT expansion 
cannot be applied directly to the scattering amplitude in systems with
more than one nucleon~\cite{Weinberg:1990rz}. He
argued that the $\chi$PT Lagrangian and counting rules should instead be used
to compute an NN (or NNN or \ldots) potential up to some fixed order, $n$,
in $\chi$PT. Such an expansion can then be examined for convergence with $n$.
The $\chi$PT potential $V$ was computed to $O(Q^3)$ in
Refs.~\cite{Ordonez:1995rz,Epelbaum:1999dj,Entem:2001cg},
and to $O(Q^4)$ in Refs.~\cite{Entem:2003ft,Epelbaum:2004fk}.
Consistent three-nucleon forces have been derived and implemented in such an approach~\cite{VanKolck:1994yi,Epelbaum:2002vt}.

However, while there is a $\chi$PT expansion for $V$, the resulting nuclear binding energies
(and other observables) contain effects to all orders in the chiral expansion: there is
no obvious perturbative expansion for them. In practice, chiral EFT for few-nucleon systems
is often implemented as described in the previous paragraph, but with the Hamiltonian acting
only on a limited space: in
momentum space a cutoff $\Lambda$ in the range $450 < \Lambda < 800$ MeV must
be imposed~\cite{Marji:2013uia}. From now on when we use the term chiral EFT in the context
of few-nucleon systems we mean calculations that are carried out in this way. A formal justification of the $Q$-expansion (e.g., via the distorted-wave Born approximation
evaluation of higher-order contributions~\cite{Long:2011xw,Long:2011qx,Valderrama:2011mv} or use of a relativistic propagator~\cite{Epelbaum:2012ua,Epelbaum:2013ij}) requires a more sophisticated power counting~\cite{Nogga:2005hy,Birse:2009my,Phillips:2013fia}. Nevertheless, in practice, the convergence of chiral EFT calculations for observables can
be examined {\it a posteriori} to
see if they inherit the $Q$-expansion that has been used for the potential.

In two recent papers, EKM estimated
the errors that arise from truncation of the chiral EFT expansion
at a finite order~\cite{Epelbaum:2014efa,Epelbaum:2014sza}
(see also Ref.~\cite{Binder:2015wpa}).
Similar prescriptions have previously been used in other EFT contexts (see, e.g., Refs.~\cite{Phillips:1999hh,Griesshammer:2012we}).
Such estimates apply to individual observables (such as the total cross
section for neutron-proton scattering at a given lab energy or nucleon electric and magnetic polarizabilities).
They are independent of procedures used to
fit LECs to two-body scattering data at each order. While Bayesian analysis could also be applied to
those procedures that is not our concern here; it will be the focus of a future publication~\cite{Wesolowski:2015}.

Instead, EKM assume that the EFT expansion holds for individual observables $X(p)$, i.e.,
\beq
X(p)=X_0\sum_{n=0}^k c_n(p) Q^n \;,
\label{eq:EKMexp}
\eeq
with $Q$ the EFT expansion parameter, and $c_1=0$ in the Weinberg expansion for NN scattering in chiral EFT.
Cumulative sums at LO, NLO, N$^2$LO, N$^3$LO, and N$^4$LO are then given by:
\begin{eqnarray}
X^{\rm LO}(p)&=&c_0(p)  \;, \\
X^{\rm NLO}(p)&=& \sum_{n=0}^2 c_n(p) Q^n  \;, \\
X^{\rm N^jLO}(p)&=&\sum_{n=0}^{j+1} c_n(p) Q^n, \; j=2, 3, 4 \;.
\label{eq:partialsums}
\end{eqnarray}

EKM also assume that the dominant error at order $k$ comes from the first 
omitted---($k+1$)$^{\rm th}$---term. Two ingredients
go into their estimate of this term. The first is to identify the EFT
expansion parameter $Q$, defined as
  \beq
    Q \equiv \max\left( \frac{p}{\Lambda_b},\ \frac{m_\pi}{\Lambda_b} \right)
    \;.
    \label{eq:Qexpansion}
  \eeq
Note that, in contrast to pQCD, this is a momentum-dependent
expansion parameter, and so the expansion will perform differently at different kinematic points.
Furthermore, to know $Q$ we must identify $\Lambda_b$, the breakdown scale of the EFT.
In Refs.~\cite{Epelbaum:2014efa,Epelbaum:2014sza}, EKM estimate $\Lambda_b$
from error plots of the fit phase shifts. The second ingredient is to determine the shift beyond N$^j$LO as:
\beq
\Delta X^{\rm N^jLO}=Q^{j+2} {\rm max}(|c_0|, |c_1|, \ldots, |c_{j+1}|) \;,
\label{eq:DeltaX}
\eeq
where the $c_n$'s are defined as above.

In Refs.~\cite{Epelbaum:2014efa,Epelbaum:2014sza} the expressions for the theory
error are defined via differences of the partial sums (\ref{eq:partialsums}), but the result may
be summarized compactly according to Eq.~\eqref{eq:DeltaX}. The similarity of this prescription
to the simplest analytic form obtained above with \Aeps$^{(1)}$ priors,
the CH procedure written in Eq.~\eqref{eq:dcoeffs},
is evident.
For a given observable, the value of $Q$ that
is identified defines the perturbative expansion parameter, and the EKM uncertainty
is the maximum coefficient times the first omitted power of $\Q$.  Up to factors
of order unity, this is what Eq.~\eqref{eq:dcoeffs} predicts
for the 68\% (``$1\sigma$'') DOB interval.  There is then clearly a semi-quantitative
correspondence.  We now make a quantitative comparison using the various
priors from Table~\ref{tab:priors}.

\begin{table}[tbh]
   \setlength{\tabcolsep}{4pt}
   \caption{Order-by-order calculations for $\signp$ in mb
    for $R= 0.9\,$fm
    from EKM~\cite{Epelbaum:2014efa,Epelbaum:2014sza,Epelbaum:2015private}.
    Lab energy $T_{\rm lab}$ and relative momentum $p_{\rm rel}$ are in units
    of MeV.  \label{tab:sigmanp_for_R0p9}
    }
  \begin{ruledtabular}
  \begin{tabular}{ccrrrrr}
 $T_{\rm lab}$ \strut & $p_{\rm rel}$  & $\sigma_{\rm LO}$ & $\sigma_{\rm NLO}$ &
    $\sigma_{\rm N^2LO}$  & $\sigma_{\rm N^3LO}$ & $\sigma_{\rm N^4LO}$ \\
    \hline
   50 & 153 & 183.6 & 166.5 & 167.0 & 166.8 & 167.5 \\
   96 & 212 &  84.8 &  75.1 &  78.3 &  77.5 &  78.0 \\
  143 & 259 &  52.5 &  49.1 &  54.2 &  53.7 &  53.9 \\
  200 & 307 &  34.9 &  35.9 &  42.6 &  43.2 &  42.7 \\
  \end{tabular}
  \end{ruledtabular}
\end{table}

\begin{table}[tbh]
   \setlength{\tabcolsep}{4pt}
   \caption{Order-by-order calculations for $\signp$ in mb
    for $R= 1.2\,$fm
    from EKM~\cite{Epelbaum:2014efa,Epelbaum:2014sza,Epelbaum:2015private}.
    Lab energy $T_{\rm lab}$ and relative momentum $p_{\rm rel}$ are in units
    of MeV. \label{tab:sigmanp_for_R1p2}
    }
  \begin{ruledtabular}
  \begin{tabular}{ccrrrrr}
 $T_{\rm lab}$ \strut & $p_{\rm rel}$  & $\sigma_{\rm LO}$ & $\sigma_{\rm NLO}$ &
    $\sigma_{\rm N^2LO}$  & $\sigma_{\rm N^3LO}$ & $\sigma_{\rm N^4LO}$ \\
    \hline
   50 & 153 & 159.4 & 164.8 & 165.6 & 167.2 & 167.9 \\
   96 & 212 &  60.2 &  68.9 &  71.3 &  78.1 &  78.5 \\
  143 & 259 &  30.8 &  38.6 &  41.4 &  52.6 &  52.7 \\
  200 & 307 &  17.2 &  22.5 &  25.0 &  38.6 &  38.3 \\
  \end{tabular}
  \end{ruledtabular}
\end{table}

\begin{table}[tbh]
   \setlength{\tabcolsep}{4pt}
   \caption{Dimensionless coefficients from the expansion of
    $\signp = \sigma_{\rm LO} \sum_{n=0}^5 c_n \Q^n$
    for $R= 0.9\,$fm from Table~\ref{tab:sigmanp_for_R0p9},
    with $\Q = p_{\rm rel}/600\,$MeV. \label{tab:sigmanp_coeffs_for_R0p9}
    }
  \begin{ruledtabular}
  \begin{tabular}{crrrrrrr}
  $T_{\rm lab}$ & $c_0$ & $c_1$ & $c_2$ & $c_3$ & $c_4$ & $c_5$  \\
    \hline
   50 & 1.0 & 0.0 & $-1.43$ & 0.16 & $-0.26$ & 3.5\phantom{4}  \\
   96 & 1.0 & 0.0 & $-0.92$ & 0.86 & $-0.61$ & 1.07 \\
  143 & 1.0 & 0.0 & $-0.35$ & 1.21 & $-0.27$ & 0.25 \\
  200 & 1.0 & 0.0 & 0.11    & 1.44 &  0.25   & $-0.41$ \\
  \end{tabular}
  \end{ruledtabular}
\end{table}

\begin{table}
   \setlength{\tabcolsep}{4pt}
   \caption{Dimensionless coefficients from the expansion of
    $\signp = \sigma_{\rm LO} \sum_{n=0}^5 c_n \Q^n$
    for $R= 1.2\,$fm from Table~\ref{tab:sigmanp_for_R1p2},
    with $\Q = p_{\rm rel}/400\,$MeV. \label{tab:sigmanp_coeffs_for_R1p2}
    }
  \begin{ruledtabular}
  \begin{tabular}{crrrrrrr}
  $T_{\rm lab}$ & $c_0$ & $c_1$ & $c_2$ & $c_3$ & $c_4$ & $c_5$  \\
    \hline
   50 & 1.0 & 0.0 & 0.23 & 0.09 & 0.47 & 0.54  \\
   96 & 1.0 & 0.0 & 0.51 & 0.27 & 1.43 & 0.16 \\
  143 & 1.0 & 0.0 & 0.60 & 0.33 & 2.07 & 0.03 \\
  200 & 1.0 & 0.0 & 0.52 & 0.32 & 2.28 & $-0.07$ \\
  \end{tabular}
  \end{ruledtabular}
\end{table}

\begin{figure}[bth]
 \includegraphics[width=0.98\columnwidth]{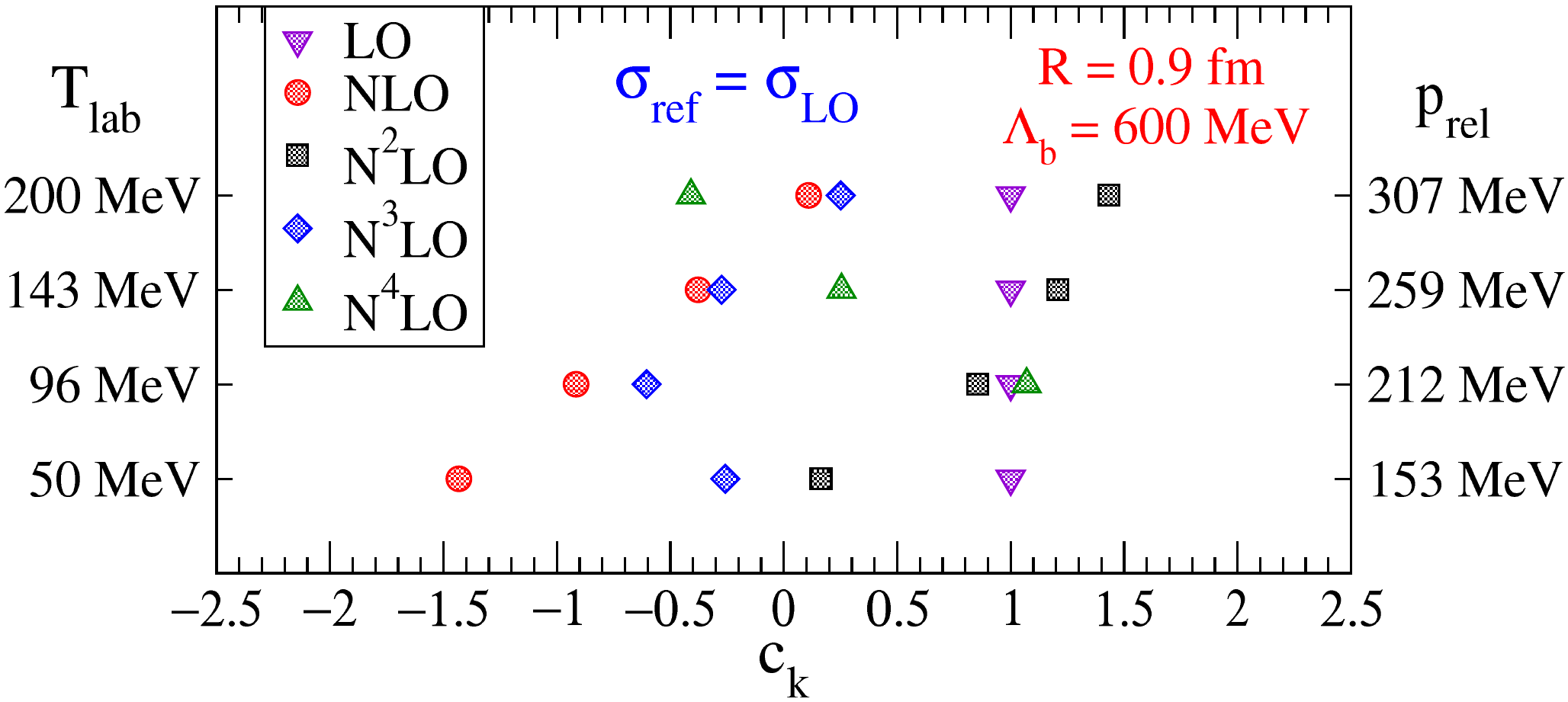}
 \caption{(color online) Chiral EFT expansion coefficients
 from Table~\ref{tab:sigmanp_coeffs_for_R0p9} for $\signp$ at four different energies
 using potentials with regulator parameter $R=0.9\,$fm and $\Lambda_b = 600\,$MeV. Note that the coefficient
 $c_5=3.5$ at $T_{\rm lab}=50$ MeV is off scale.}
 \label{fig:scaledR0p9Lam600}
\end{figure}

\begin{figure}[bth]
 \includegraphics[width=0.98\columnwidth]{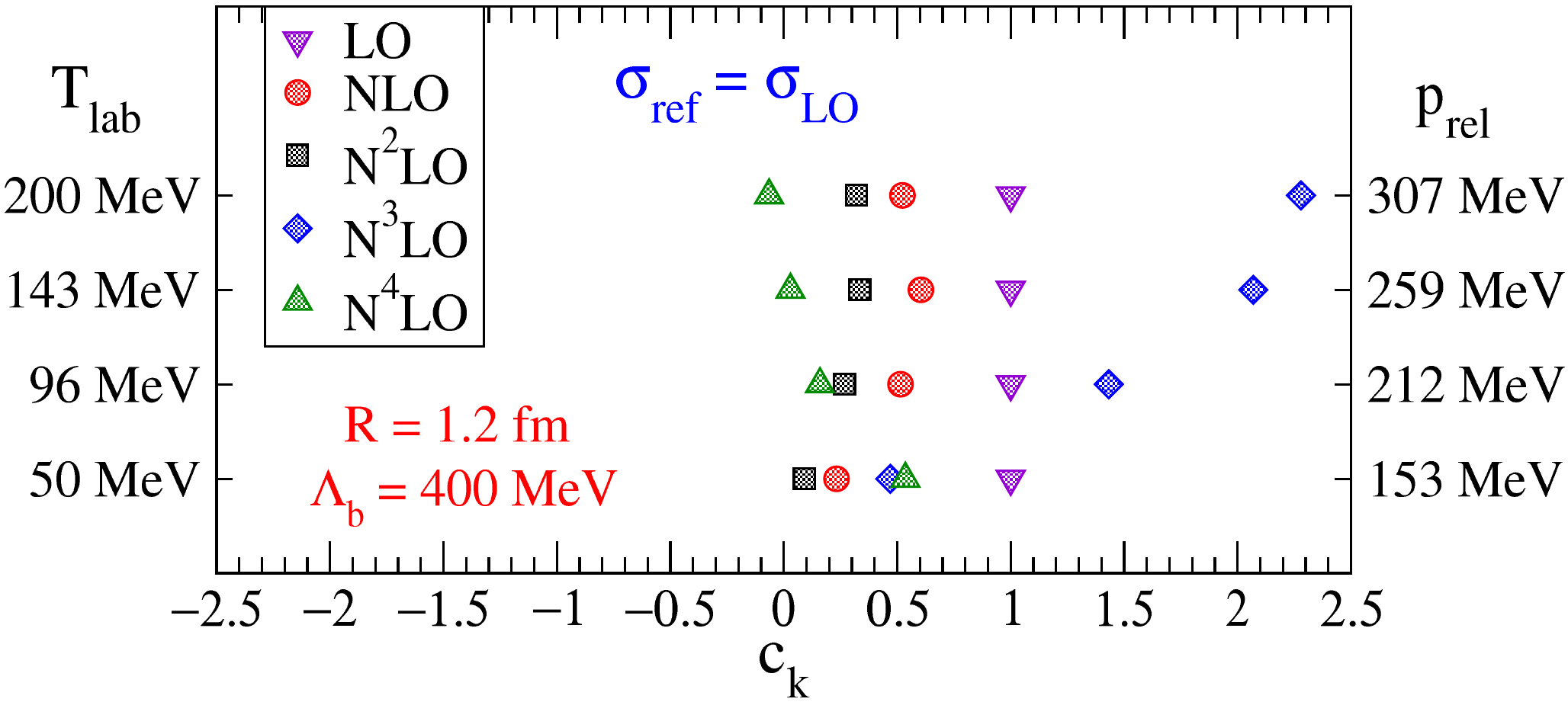}
 \caption{(color online) Chiral EFT expansion coefficients from Table~\ref{tab:sigmanp_coeffs_for_R0p9}
 for $\signp$ at four different energies
 using potentials with regulator parameter $R=1.2\,$fm and $\Lambda_b = 400\,$MeV.
 }
 \label{fig:scaledR1p2Lam400}
\end{figure}

We note that in chiral EFT, having posited a $\Q$ expansion,
we do not find the coefficients directly but extract them
from the calculations at different orders. In contrast, for the QCD expansions, the coefficients
are calculated independently of each other.
Thus the EFT application will require additional empirical verification, see Sec.~\ref{sec:scalechecking}.

\subsection{The pattern of EFT coefficients in EKM's result}

In Refs.~\cite{Epelbaum:2014efa} and \cite{Epelbaum:2015private}, results for
the neutron-proton total cross section at various energies are given for
several values of a coordinate-space regulator parameter $R$.  These  provide
empirical tests of the priors from Table~\ref{tab:priors}.
The order-by-order cross sections are given in Tables~\ref{tab:sigmanp_for_R0p9}
and \ref{tab:sigmanp_for_R1p2} for $R=0.9\,$fm and $R=1.2\,$fm, respectively.
In line with Eq.~\eqref{eq:obsexp}, 
we write the cross section at order $k$ in the chiral EFT expansion as
\beq
  \signp(\Elab) \approx \sigmaref \sum_{n=0}^{k} c_n
    \left(\frac{p}{\Lambda_b}\right)^n
    \;,
    \label{eq:sigmaexp}
\eeq
where $\sigmaref$ is a reference cross section that might be taken as
$\sigma_{\rm LO}$, as we do here, or the \NNNNLO\ result, or
the experimental value.  The analysis is not sensitive to this choice.
The breakdown scale $\Lambda_b$ was identified in Ref.~\cite{Epelbaum:2014efa}
as $\Lambda_b = 600\,$MeV for cutoffs $R = 0.8$, 0.9,
and 1.0\,fm, $\Lambda_b = 500\,$MeV for $R =1.1\,$fm, and $\Lambda_b = 400\,$MeV
for $R = 1.2\,$fm.  Note that this decrease in $\Lambda_b$ with increasing $R$
corresponds to the change in the regulator cutoff scale rather than a change in
the intrinsic underlying breakdown scale.
In Ref.~\cite{Furnstahl:2014jpg} it was emphasized that residuals
for a $k$th-order EFT calculation had two
types of errors: regulator artifacts dictated by the imposed cutoff
$\Lambda$ ($\sim 1/R$ in this context) and truncation errors in the Hamiltonian dictated by the underlying
breakdown scale $\Lambda_b$.  EKM\ do not make this distinction
in their notation, i.e., they use $\Lambda_b$ for both.

Under the EKM choices for $\Lambda_b$, the dimensionless $c_n$ coefficients
are given in Tables~\ref{tab:sigmanp_coeffs_for_R0p9} and
\ref{tab:sigmanp_coeffs_for_R1p2} and Figs.~\ref{fig:scaledR0p9Lam600}
and \ref{fig:scaledR1p2Lam400} for the R=0.9 fm and R=1.2 fm cases respectively.
Although the coefficients in both cases are natural, rather different patterns are seen.
As discussed and illustrated by EKM
(e.g., see Fig.~2 of Ref.~\cite{Epelbaum:2014sza}), the softer cutoff shifts
contributions between different chiral orders so that the systematic pattern
of corrections is disrupted.  In particular, corrections at orders \NNLO\ and
\NNNNLO, which are purely from non-analytic terms in the chiral expansion, become
heavily regulated by the soft cutoff.
This has the consequence
that the corresponding coefficients are anomalously small---which may, in turn result in  \NNNLO\ coefficients being somewhat large.
This pattern is seen in Fig.~\ref{fig:scaledR1p2Lam400}, but what is shown there
is insufficient to definitively
establish there is an inter-order correlation due to regulator artifacts. Here we 
focus on the $R=0.9\,$fm example to ensure that the pattern is
primarily driven by the inheritance of
naturalness from the fit low-energy constants (LECs), and not by regulator artifacts that spring from a choice
of $R$ that makes the $\Lambda \sim 1/R$ effects predominate over the physics at
$\Lambda_b$ that was integrated out of the theory.

\subsection{DOB intervals from a Bayesian analysis}

There is a minimum of necessary information that must exist between the prior
and data in order for
the resulting posterior to accurately describe the above distributions.
Two extremes exist: a large supply of precise and accurate data paired
with an uninformative prior (or, even worse, an informative yet incorrect prior) and
a small amount of data paired with a precisely and accurately defined
prior. Each of these situations may result in realistic posteriors as lack of
information in one realm is compensated by abundance in the other.  In practice,
though, we conduct analyses between these extremes.  We are often able to define a
reasonable, and appropriately loose, prior that is subsequently
fine-tuned by a modest amount of data. We will now show that each
of the priors defined in Table \ref{tab:priors} may be considered reasonable
representations of naturalness in the EFT-coefficient distribution obtained in the
previous subsection. The DOB intervals that result from Bayesian analyses
using these priors show agreement and increased similarity at low $\Q$ and high
$k$---where the strength of available data is greatest.

\begin{table*}[tbh!]
\setlength{\tabcolsep}{9pt}
\caption{DOB intervals for EKM $\sigma_{np}$ with $R=0.9\,$fm using prior Set $\Aepsone$.}
\label{tab:EKMresults}
  \begin{tabular}{c|c|c|c|c|c|c|c|c|}
 DOB &  $\mbox{E}_{\rm lab}$ & $\Q$ & $\mbox{LO}'$ &$\mbox{LO}$& $\mbox{NLO}$ & $\mbox{\NNLO}$ &  $\mbox{\NNNLO}$ &  $\mbox{\NNNNLO}$ \\
  \hline
 68\%      & \multirow{2}{*}{50} & \multirow{2}{*}{0.255} &  0.40 & 0.102 & 0.024 & 0.0055 & 0.0013 & 0.00079 \\
$\times 183.6$\,mb & &       &  73. & 19.  & 4.4   & 1.0  & 0.24  & 0.15  \\
  \hline
 68\%    & \multirow{2}{*}{96} & \multirow{2}{*}{0.354} &  0.55 & 0.20 & 0.045 & 0.0142 & 0.0047 & 0.0017 \\
$\times 84.8$\,mb  & &       &  47. & 17.  & 3.8   & 1.2  & 0.40  & 0.15  \\
\hline
68\%   & \multirow{2}{*}{143}   &  \multirow{2}{*}{0.432}     &    0.68 & 0.29 & 0.082 & 0.038 &  0.015  & 0.0064 \\
 $\times 52.5$\,mb & &       &  35. &  15. &  4.3  & 2.0  & 0.81  & 0.34  \\
\hline
68\%   &  \multirow{2}{*}{200}    &  \multirow{2}{*}{0.511}     &    0.80 & 0.41 &  0.136 & 0.089 &  0.043 &  0.021 \\
 $\times 34.9$\,mb  &  &       &  28. & 14.  &  4.7  & 3.1  &  1.5 & 0.73  \\
\hline\hline
95\%       & \multirow{2}{*}{50}   & \multirow{2}{*}{0.255}    &   2.6 &  0.650 & 0.061 & 0.0103 &  0.0022 & 0.0012   \\
$\times 183.6$\,mb & &       &  470 & 120  & 11.   & 1.9  & 0.40  & 0.23  \\
\hline
95\%      &  \multirow{2}{*}{96}  &  \multirow{2}{*}{0.354}     &   3.5 &   1.25 &  0.115 &  0.027 & 0.0079 & 0.0027   \\
$\times 84.8$\,mb  & &       &  300 & 110  & 9.8   & 2.3  & 0.67  & 0.23  \\
\hline
95\%     & \multirow{2}{*}{143}   &  \multirow{2}{*}{0.432}     &   4.3 &  1.87 &  0.21 & 0.072 &  0.026 &  0.010 \\
 $\times 52.5$\,mb & &       &  230 & 98.  &  11.  & 3.8  & 1.4  &  0.53 \\
  \hline
95\%     &  \multirow{2}{*}{200}    &   \multirow{2}{*}{0.511}    &    5.1 &  2.6  &  0.35 & 0.17  &  0.071 &  0.033 \\
 $\times 34.9$\,mb  &  &       &  180 & 91.  &  12.  & 5.9  &  2.5 & 1.1  \\
    \hline
  \end{tabular}
\end{table*}

\begin{table*}[tbh!]
 \setlength{\tabcolsep}{9pt}
 \caption{DOB intervals for EKM $\sigma_{np}$ scaled by $\sigma_{\rm LO}$
with $R=0.9\,$fm.  Results for prior Sets \Aeps, B (with $\sigma = 1.0$), and \Ceps, all without the leading-omitted-term approximation.}
 \label{tab:EKMresults_setA_approx1}
 \begin{tabular}{c|c|c|c|c|c|c|c|c|c|}
  & set &  $\mbox{E}_{\rm lab}$ & $\Q$ & $\mbox{LO}'$ &$\mbox{LO}$& $\mbox{NLO}$ & $\mbox{\NNLO}$ &  $\mbox{\NNNLO}$ &  $\mbox{\NNNNLO}$  \\
  \hline
  & \Aeps   & &  & 0.43  & 0.11  & 0.025  & 0.0055  & 0.0013  & 0.00080 \\
  &\Ceps &    50 & 0.255 & 0.48  & 0.12  & 0.028  & 0.0053  & 0.0011  & 0.00056 \\
   &B &    &  &  0.29 & 0.073 & 0.022 & 0.0052 & 0.0013 & 0.00076   \\
  \cline{2-10}
  &  \Aeps   &  &  & 0.59  & 0.21  & 0.048  & 0.015   & 0.0048  & 0.0018  \\
   &\Ceps &     96 & 0.354 & 0.69  & 0.25  & 0.060  & 0.019   & 0.0058  & 0.0021  \\
  \multirow{2}{*}{68\%}& B &   &      & 0.40 & 0.143 & 0.043 & 0.014 & 0.0047 & 0.0017 \\
  \cline{2-10}
  & \Aeps & &  & 0.74  & 0.32  & 0.089  & 0.040   & 0.016   & 0.0067  \\
  & \Ceps &   143 & 0.432 & 0.87  & 0.38  & 0.088  & 0.043   & 0.015   & 0.0059  \\
   & B  &   &    & 0.51 & 0.22 & 0.080 & 0.038 & 0.016 & 0.0065  \\
 \cline{2-10}
  & \Aeps    & & & 0.91  & 0.46  & 0.15   & 0.097   & 0.046   & 0.022   \\
  &\Ceps &   200 & 0.511 & 1.08  & 0.58  & 0.14   & 0.096   & 0.041   & 0.019   \\
   & B  &    &        & 0.63 & 0.32 & 0.14 & 0.091 & 0.044 & 0.022 \\
 \hline\hline
   & \Aeps   &    &      &  2.7  & 0.69  & 0.066  & 0.011   & 0.0023  & 0.0013  \\
  &\Ceps &   50   &    0.255   &  3.3  & 0.85  & 0.089    & 0.014   & 0.0027  & 0.0013  \\
  & B &  &      &   0.67 & 0.172 & 0.042 & 0.0091 &  0.0021 & 0.0012 \\
  \cline{2-10}
   &  \Aeps   &   &     &  3.8  &  1.3  & 0.13   & 0.030   & 0.0088  & 0.0030  \\
   &\Ceps &     96  &   0.354    &  4.8  &  1.7  & 0.20   & 0.050   & 0.0142  & 0.0049  \\
  \multirow{2}{*}{95\%}  & B &    &     & 0.97 & 0.34 & 0.088 & 0.026 & 0.0083  & 0.0029 \\
  \cline{2-10}
   &  \Aeps   &   &      & 4.7   & 2.0   & 0.24   & 0.083   & 0.030   & 0.012   \\
    & \Ceps &   143   &   0.432    & 6.0   & 2.6   & 0.29   & 0.114   & 0.038   & 0.014   \\
    & B &  &        & 1.22  & 0.53 & 0.17 & 0.071 & 0.028  & 0.0115 \\
  \cline{2-10}
    & \Aeps  &   &      & 5.7   & 2.9   & 0.41   & 0.20    & 0.088   & 0.041   \\
  & \Ceps &   200   &   0.511    & 7.4   & 3.8   & 0.47   & 0.26    & 0.100   & 0.043   \\
    & B &   &      & 1.53  & 0.78 & 0.29 & 0.173 & 0.081 & 0.040  \\
   \hline
 \end{tabular}
\end{table*}

\begin{figure}[b!]
  \includegraphics[width=0.95\columnwidth]{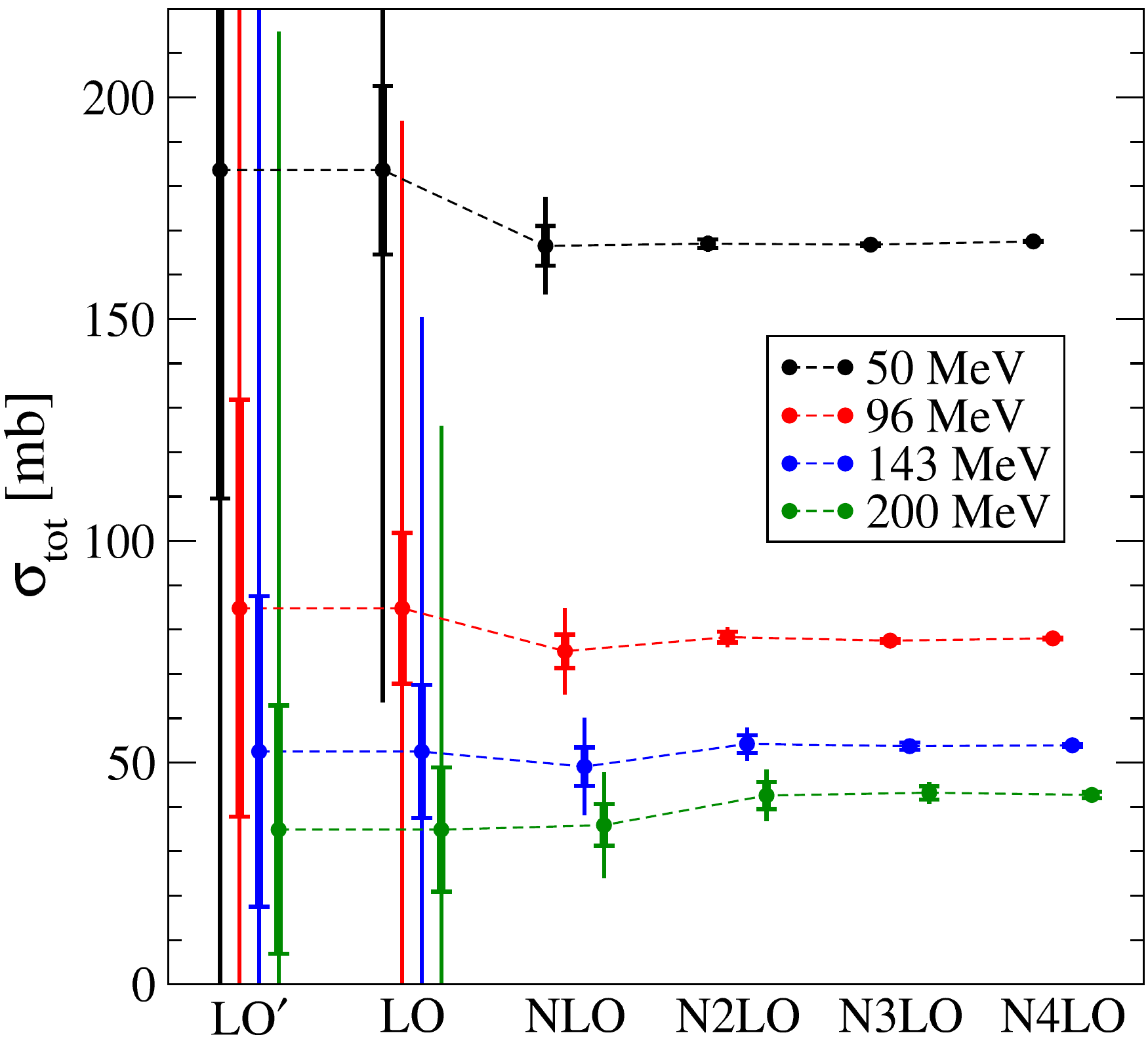}
  \caption{(color online) Cross sections at different energies and orders from EKM, with
  DOB intervals at each order using Set \Aeps\ priors.
  The thick error bars indicate 68\% DOB intervals while the thin
  error bars indicate 95\% DOB intervals.}
  \label{fig:EKM_Tlab_together_setA_eps_approx0}
\end{figure}

\begin{figure*}[tbp]
  \includegraphics[width=0.85\columnwidth]{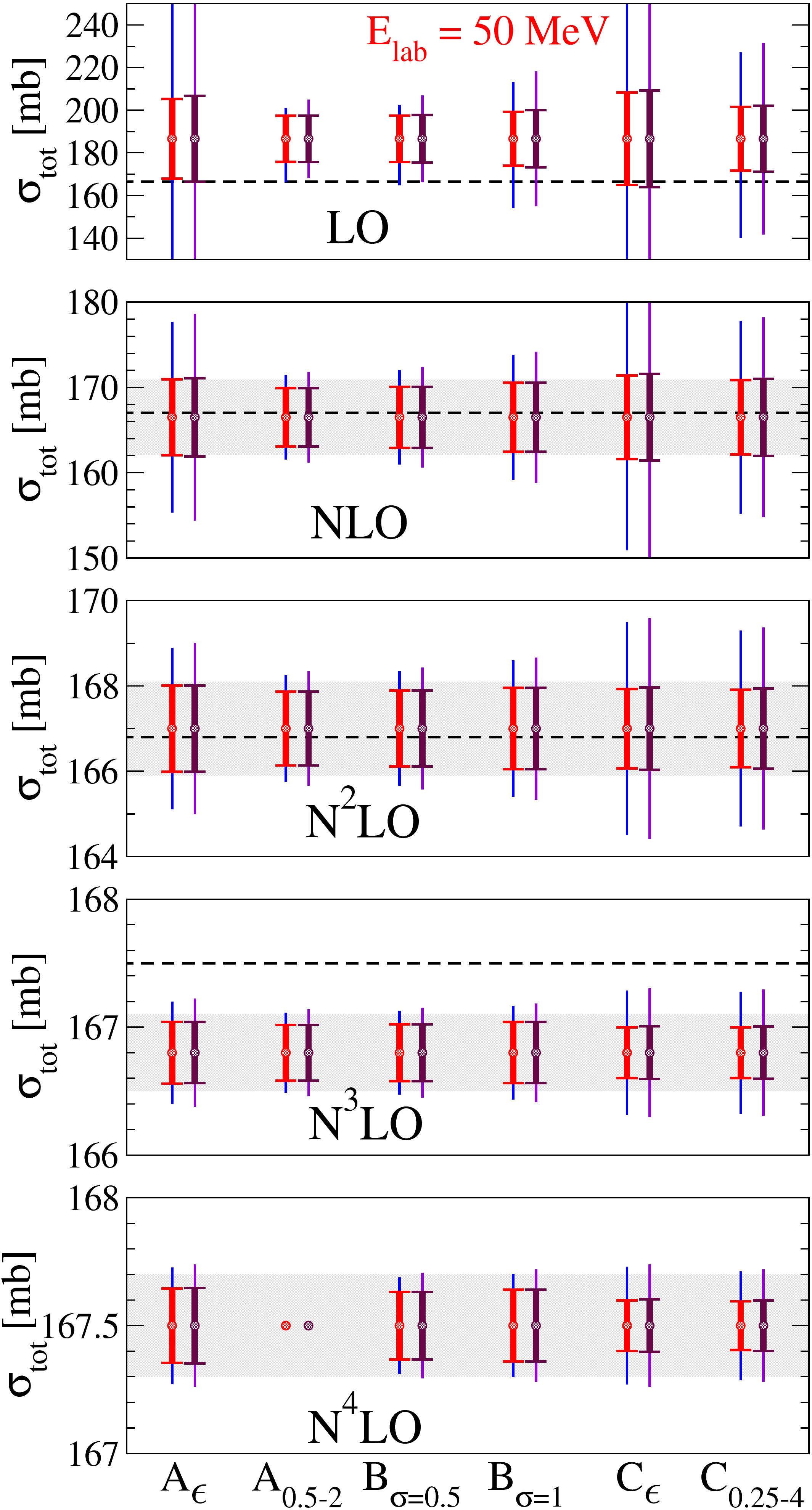}
  \hspace*{.3in}
  \includegraphics[width=0.85\columnwidth]{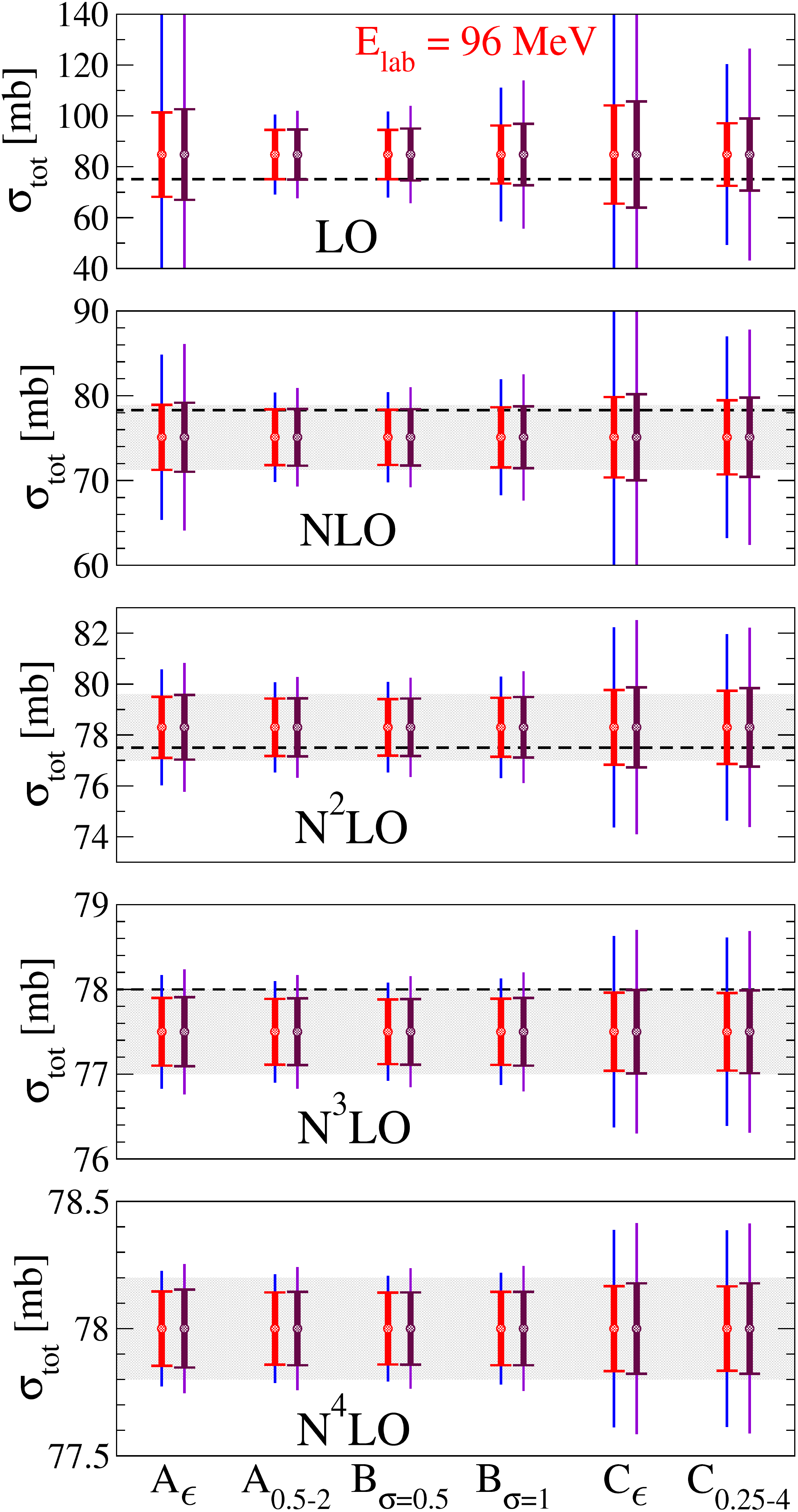}
  \caption{(color online) Cross sections at 50\,MeV and  96\,MeV for all orders from EKM, with
  DOB intervals at each order using a wide variety of prior sets. Note the change
  in scale at each order. 
  The thick error bars indicate 68\% DOB intervals while the thin
  error bars indicate 95\% DOB intervals. In each panel the dashed line is the result of the next-order calculation
  (NLO at LO, N$^2$LO at NLO, etc.), shown to facilitate an assessment of the statistical consistency of
  different prior choices. 
   For each prior choice, the intervals on
  the left are from keeping only the first omitted term while those on the
  right are including four omitted terms.
  The shaded bands indicate the uncertainty from EKM.
  }
  \label{fig:EKM_Tlab_96_all_orders_all_sets}
\end{figure*}

\begin{figure*}[tbp]
  \includegraphics[width=0.85\columnwidth]{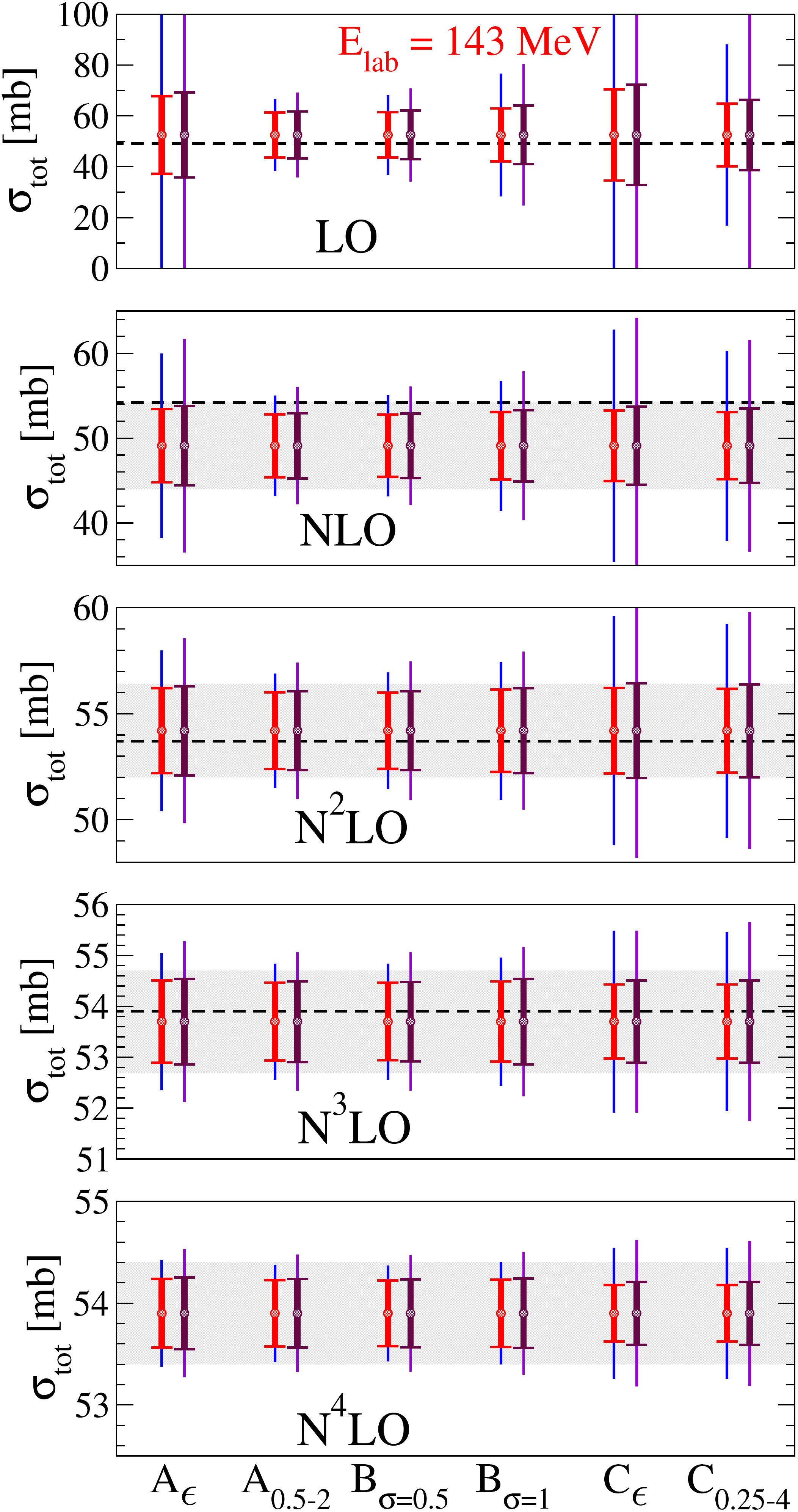}
  \hspace*{.3in}
  \includegraphics[width=0.85\columnwidth]{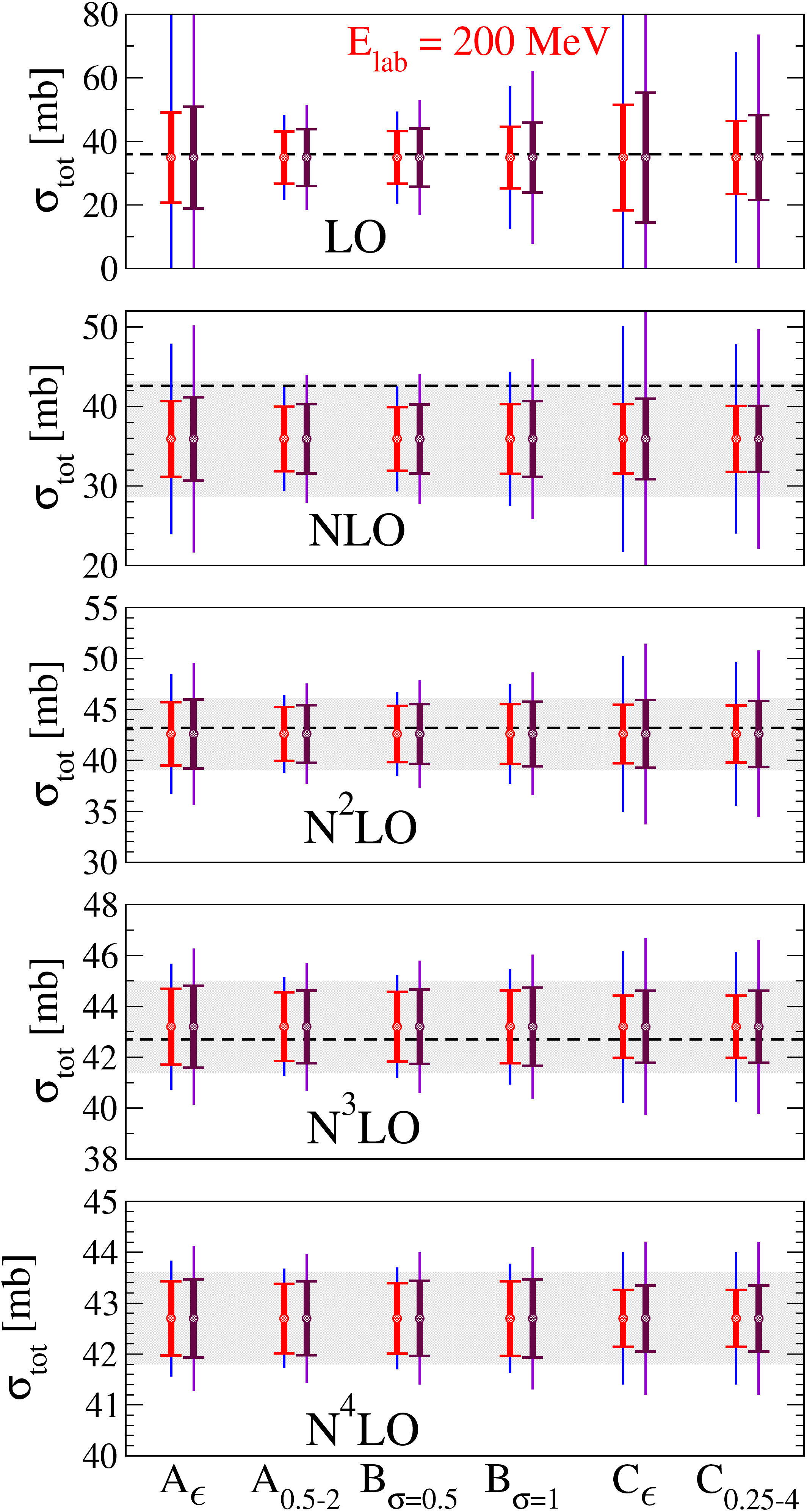}
  \caption{(color online) Cross sections at 143\,MeV and 200\,MeV for all orders from EKM, with
  DOB intervals at each order using a wide variety of sets. Note the change
  in scale at each order.The thick error bars indicate 68\% DOB intervals while the thin
  error bars indicate 95\% DOB intervals. In each panel the dashed line is the result of the next-order calculation
  (NLO at LO, N$^2$LO at NLO, etc.), shown to facilitate an assessment of the statistical consistency of
  different prior choices. 
   For each prior choice, the intervals on
  the left are from keeping only the first omitted term while those on the
  right are including four omitted terms.
  The shaded bands indicate the uncertainty from EKM.
  }
  \label{fig:EKM_Tlab_200_all_orders_all_sets}
\end{figure*}

In Fig.~\ref{fig:EKM_Tlab_together_setA_eps_approx0}, cross sections
from Table~\ref{tab:sigmanp_for_R0p9} for
$R = 0.9\,$fm at four
different energies are plotted order-by-order in the chiral expansion,
with error bars indicating the 68\% and 95\% DOB intervals if we
adopt prior set \Aeps.
The $\mbox{LO}'$ error bars are from the calculation for the posterior
of $\Delta_0$ while the $\mbox{LO}$ error bars are from the posterior of
$\Delta_1$. When calculating $\Delta_1$ we have $k=1$ and $n_c=1$,
so the resulting error bar is simply $\Q$ times the $\mbox{LO}'$ one.
This is the correct error estimate for a LO chiral
EFT calculation of NN scattering, as long as we know {\it a priori} that the coefficient
$c_1$ in the expansion (\ref{eq:sigmaexp}) is identically zero.

Cross sections at subsequent orders generally fall within the DOB
intervals of lower-order error analyses---in accord with the DOB intervals' statistical
interpretation.  The order-by-order decrease in the error bars primarily
reflects the additional factors of $\Q$ with each successive order.
The very conservative assumption for $\pdf(\cbar)$ used here, which encodes ignorance
of its scale even though we anticipate naturalness, leads to long tails in the
posterior for the lowest orders and correspondingly large 95\% DOB intervals---
further reflecting the non-Gaussian nature of this distribution.
When we use a form for $\pdf(\cbar)$ that reflects the expectation of naturalness,
the long tails are suppressed and the 95\% DOB intervals are closer to
``$2\sigma$'' errors, although the pdfs do remain non-Gaussian in general.

Table \ref{tab:EKMresults} is an analytical compilation of DOB intervals in
the limiting case of $\Aeps$ with the leading-term approximation,
see Eq.~(\ref{eq:dcoeffs}).  The top number in each cell has been calculated
using coefficients of Table~\ref{tab:sigmanp_coeffs_for_R0p9}, which are scaled
with the $\sigmaref$ corresponding to each energy.
The lower number is the resulting DOB interval in units of mb with the factor of $\sigmaref=\sigma_{\rm LO}$ included.
The 95\% DOB interval being more than 6 times broader than the corresponding
68\% DOB interval at $\mbox{LO}'$ and $\mbox{LO}$ emphasizes the strength of tails within these posteriors.

Representative numerical results for the various prior sets are given in
Table~\ref{tab:EKMresults_setA_approx1}. Though we omit mention of $\sigmaref$,
values here should also be multiplied by the energy-appropriate $\sigmaref$ (i.e., $\sigma_{\rm LO}$) to
obtain DOB values in units of mb.  Systematically, we observe that the ratios of DOB intervals
between prior sets are the same across all 4 energies for $\mbox{LO}'$ and $\mbox{LO}$ as all $c_1$'s are
0 and we have scaled all $c_0$'s to the value 1.0.  Thus, given the same set of
coefficients, all posteriors scale similarly with energy.
Table~\ref{tab:EKMresults_setA_approx1} also shows that the ensemble prior in
Set~$\Ceps$ generally predicts 68\% DOB intervals quite similar to those from
Set~$\Aeps$, with much greater variation for 95\% DOB intervals for the lower orders.
From this, we see that prior choice affects the structure of the tails
more significantly than the structure of the peak. This is indicative of the strength of information coming
from the data and the prior at different points in the distribution.
Though Set B results in significantly narrower DOB intervals at low $k$,
the EFT coefficients provide enough information for $k \geq 2$ to modify these posteriors into
agreement with those of Sets~$\Ceps$ and~$\Aeps$.

A comparison of Set~A results in
Table~\ref{tab:EKMresults_setA_approx1} with those in Table~\ref{tab:EKMresults}
shows that the approximation of keeping only the leading omitted term is
excellent  for the 68\% DOB for $k>2$ and still quite good
for $k=1$ (which is the true leading order).  This approximation always underestimates
the interval from including higher-order terms and worsens as the expansion
parameter $\Q$ increases.  Figures~\ref{fig:EKM_Tlab_96_all_orders_all_sets} and
\ref{fig:EKM_Tlab_200_all_orders_all_sets} show that this result is general and
that the approximation is better for the 95\% interval with a less
conservative prior for $\cbar$.
One outlier is the $k=5$ prediction
at $T_{\rm lab} = 50\,$MeV where we see consequences of a $c_5$ coefficient
known to have an anomalously large value, which is an artifact of the
fitting procedure~\cite{Epelbaum:2015private}.
Note that this results in the omission of the DOB interval for $\NNNNLO$
at 50 MeV with Set $A_{0.5-2}$ as $\cbarmax$ is then less than $\cbark$, so the distribution 
is not defined in this case. 

Overall, the prior sets $\Aeps$ and $\Ceps$ appear to be too conservative for
predictions at LO; we know that $\Aeps$ and $\Ceps$ have
incorporated less information than the alternatives so it is no surprise that
their posteriors are more widely distributed.
Importantly, we find that the posteriors for $\Delta_k$ for $k\geq 2$ are
largely insensitive to the choice of prior, even for the 95\%
DOB interval.  As posteriors retain artifacts of the prior in inverse proportion
to the strength of the data, this similarity suggests that the data is
sufficiently informative that any reasonable prior is properly subservient
and thus able to adapt to evidence of the real world presented by the data.

\section{Choice of expansion parameter}

\label{sec:scalechecking}

In the previous section, the scale $\Lambda_b$ in the expansion parameter was
taken from Ref.~\cite{Epelbaum:2014efa}, where it
was extracted from error plots after the fit of the LECs.
This identification was certainly not rigorous in any statistical sense.
Therefore here we explore how $\Lambda_b$ can be extracted
from the convergence pattern of the EFT for observables.

In the case of pQCD,
Cacciari and Houdeau discussed using an expansion parameter
that is different from $\alpha_s$. They introduced a scale factor $\lambda$, so that the expansion
is in powers of $\alpha_s/\lambda$~\cite{Cacciari:2011ze}. This changes the expressions
for $\pr(\Delta_k|c_0 \ldots, c_k)$  by a rescaling of the
expansion parameter $\Q$ and a corresponding rescaling of the coefficients
themselves.
We can rewrite the series for an observable $X$ in terms of the rescaled expansion parameter and coefficients
as
\beq
  X = X_0 \sum_{n=0}^\infty (c_n \lambda^n) \times  \left(\frac{ \Q}{\lambda}\right)^n \; .
  \label{eq:lambda-tune-exp}
\eeq
In an EFT expansion this is equivalent to a rescaling of $\Lambda_b$ by a factor $\lambda$. 

Subsequent papers explored procedures for determining the value
of $\lambda$ based on various criteria:
\bi
  \I In Refs.~\cite{Bagnaschi:2014eva,Bagnaschi:2014wea}, $\lambda$ was chosen
  empirically by comparing the consistency of the computed
  DOB intervals with known higher-order calculations.
  An extra factor of $(n-1)!$ was also introduced along with $\lambda$ in
  Eq.~\eqref{eq:lambda-tune-exp}---motivated by effects
  from renormalon chains at higher orders in the expansion. The authors
  denoted the resulting scheme \CHbar.
  We have no evidence for such a factorial in our EFT expansions and
  do not consider it further here.

  \I In Ref.~\cite{Forte:2013mda}, it was proposed that with the best expansion
  parameter, the coefficients should form a normal distribution of mean $\mu$ and standard deviation $\sigma$. 
  This criterion was used to choose a value of $\lambda$. This approach is consistent
  with naturalness for the $\{c_n\}$, as long as $\mu$ and $\sigma$ are both $\mathcal{O}(1)$.
\ei

Here we explore these procedures for tuning the expansion parameter in the
EKM cross sections, and we also suggest another criterion for assessing
$\lambda$ based on the assumption of naturalness in the EFT expansion for a
particular value of $\Lambda_b$. If a $\lambda$ emerges from such
analyses that is measurably different from one, it suggests that the true
breakdown scale of the EFT expansion is not $\Lambda_b$, but instead
$\Lambda_b \lambda$.
Given the limited
number of coefficients (20 at most) at our disposal from the EKM analysis, any statistical
procedure can only determine $\lambda$, and hence $\Lambda_b$, within sizable error bars.
Our goal in this section is less to determine $\Lambda_b$, than to
establish whether the choice $\Lambda_b=600$ MeV is consistent
with our other {\it a priori} assumptions and deductions about the convergence properties of the EFT.

\subsection{Consistency checks based on higher-order calculations}

In Refs.~\cite{Bagnaschi:2014eva,Bagnaschi:2014wea}
$\lambda$ was determined by checking the consistency of \CHbar\ DOB intervals obtained with
expansion parameters $\alpha_s/\lambda$ in several large sets of pQCD
observables.
This is done by examining actual vs.\ expected success rates of the pQCD calculations.
As stated in Ref.~\cite{Jenniches}: ``For a finite set of observables
and a given model (with fixed parameters) at order $k$, the success rate is defined as the number
of observables whose subsequent-order contributions are within the uncertainty interval predicted
by the model.''

We want to use the observed success rates $n(p)/N$ for our EFT calculation to 
infer the likelihood that $p$ is the true success rate---for many different 
choices of $p$. 
If each observable being considered is uncorrelated, the success
rate should follow a binomial distribution. Therefore the likelihood for $n$ successes amongst $N$ observables, given $p$, is
\beq
	\pr(n|p,N) = \frac{N!}{n!(N-n)!} p^n (1-p)^{N-n} \;.
	\label{eq:binomial}
\eeq
We generalize the pdf (\ref{eq:binomial})
to its continuous version, the $\beta$-distribution:
\beq
	\pr(a,b|p,N) = \frac{(a+b-1)!}{(a-1)!(b-1)!}p^{a-1}(1-p)^{b-1} \;,
	\label{eq:beta}
\eeq
with $a = n+1$ and $b=N-n+1$. We can then compute confidence intervals (CIs) on $n$ (or, equivalently $a$) for a given value of $p$ (in practice we will consider only
the 68\% and 95\% CIs). This can be done using standard integrals, and the result expressed in terms
of a range of success rates that are consistent with the chosen value of $p$.

As in Refs.~\cite{Bagnaschi:2014eva,Bagnaschi:2014wea}, we have calculations of the cross sections at several orders and energies
and are trying to determine
 values of $\lambda$ that result in consistency between assumed values of $p$ and the resulting success rates $n$.
To do this, we take the set of 16 observables we have from the EKM results: calculations at LO, NLO, N$^2$LO, N$^3$LO for four different lab energies.
(Note that each observable must have a higher-order result to which it can be compared.) We then pick
  a value of $\lambda$ and proceed
 to assess the consistency of the success rates of the theory predictions for that $\lambda$ via this algorithm (adapted from pQCD to EFT for our purposes):
\be
  \I Select a grid of $p\%$ DOB intervals with $p$ ranging from 0 to 100.

    \I Use the formalism laid out in Sections~\ref{sec:CH} and \ref{sec:EKM} to compute the $p\%$ DOB interval for each observable in the set.

  \I For each next-order calculation that is within the DOB
  interval of the previous order, count one success.

  \I Take the number of successes and divide by the total number
  of observables to get the actual success rate.

  \I Compare the actual success rate for this value of $p$ with the 68\% and 95\% confidence intervals for the number of successes if $p$ were the true success rate, as computed from the distribution (\ref{eq:beta}).
\ee
This algorithm generates a function of $p$ for this value of $\lambda$. If the curve is within the $68\%$ CI
for the entire range of $p$ values we say that the value of $\lambda$ is consistent at $1\sigma$ with the performance of the perturbative series.  Moderate fluctuations outside the $1\sigma$ band over limited regions 
 of the entire $p$ domain  can indicate a statistically consistent choice for $\lambda$, but the concern is with curves that end up systematically outside the $1\sigma$ region. This can occur in one of two ways. 
If the curve starts to veer above the $1\sigma$ region then that indicates the EFT predictions are {\it too} successful.
The expansion parameter is  overestimated, which means the EFT breakdown scale is underestimated. Alternatively, the function $n(p)/N$ may deviate well below the 68\% CI, in which
case the EFT is under-performing compared to statistical expectations. In that case the stated expansion parameter is too small, i.e., $\Lambda_b$ is overestimated.
We note that this interpretation is somewhat specific to EFT: in a case where we were confident of the expansion parameter in the series we could instead use this diagnostic to probe
whether different prior choices are too conservative or too aggressive.

Here though, we try to draw conclusions on the performance of the EFT expansion that are invariant under the choice of priors defined above. We thus implement this procedure for two different prior assumptions
on $\cbar$ and the coefficients $\{c_n\}$.
In each case we use the approximation that
the leading term dominates for computational ease.
The curves do not change substantially if we go beyond the first-omitted-term approximation.

\begin{figure}
  \includegraphics[width=0.48\textwidth]{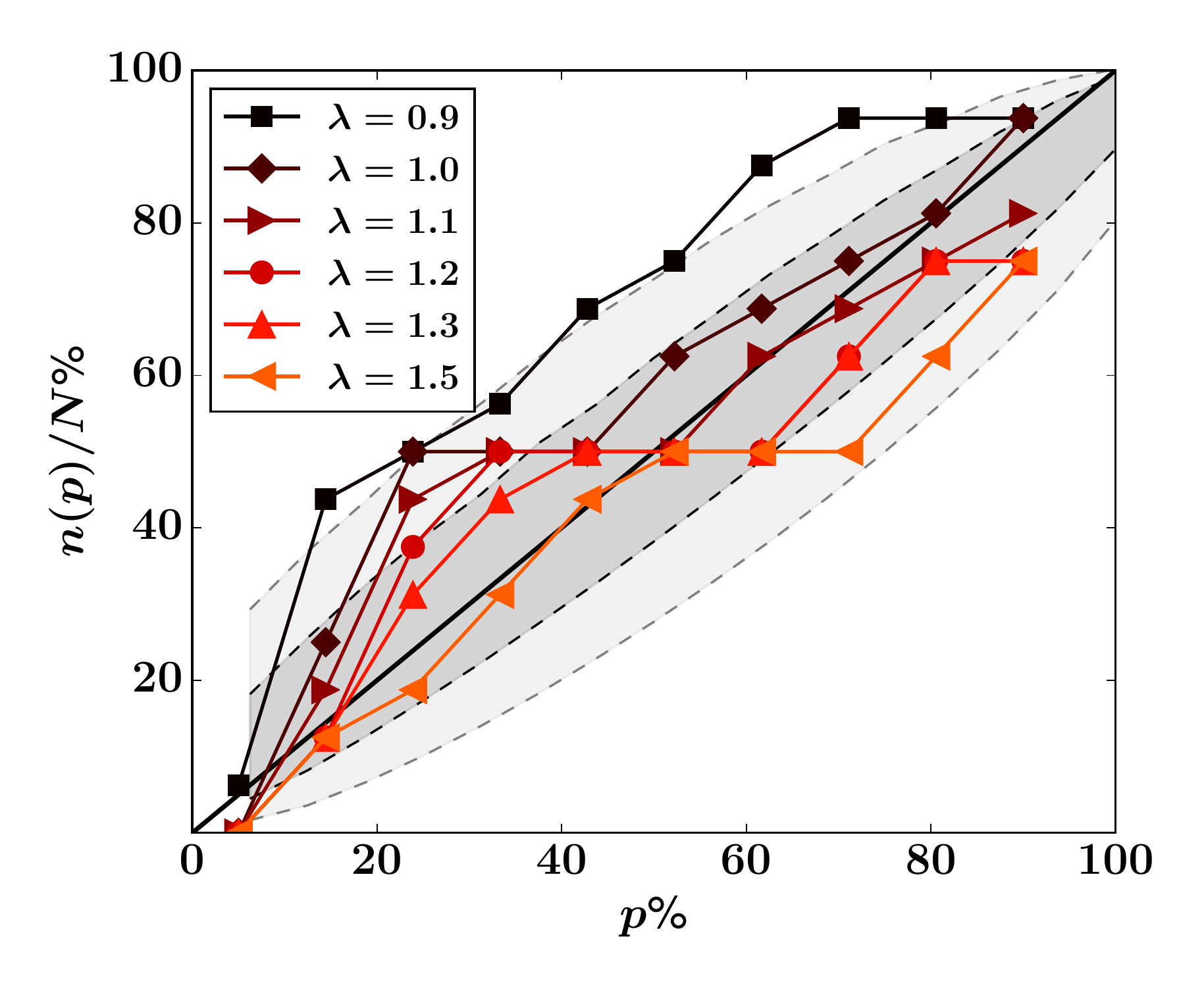}
  \caption{(color online) Empirical determination of $\lambda$
  by comparing results at different orders. The cross sections used are
  the computations with the $R=0.9\,$fm regulator. Priors are Set A$_\epsilon^{(1)}$. For full explanation see text. \label{fig:lam-success-plot-A}}
\end{figure}

We first consider Set $\Aepsone$.
The results of computing $p\%$ success rates for various values of $\lambda$ are
shown by the lines in Fig.~\ref{fig:lam-success-plot-A}.
We include $68\%$ and $95\%$ confidence
bands to evaluate which $\lambda$ curves meet our consistency
criterion.
With only 16 observables, the confidence bands are fairly wide, but still the only
curve which falls completely within the $68\%$ interval is $\lambda = 1.3$. The original
expansion parameter at $\lambda = 1$ spends some time above the $1\sigma$ region, which may reflect that DOB intervals resulting from this prior are too conservative; i.e., the actual success rate
regularly exceeds the DOB that has been assigned. This is consistent with our
earlier observation that Set $\Aeps$ priors produce overly conservative DOB intervals.

\begin{figure}
  \includegraphics[width=0.48\textwidth]{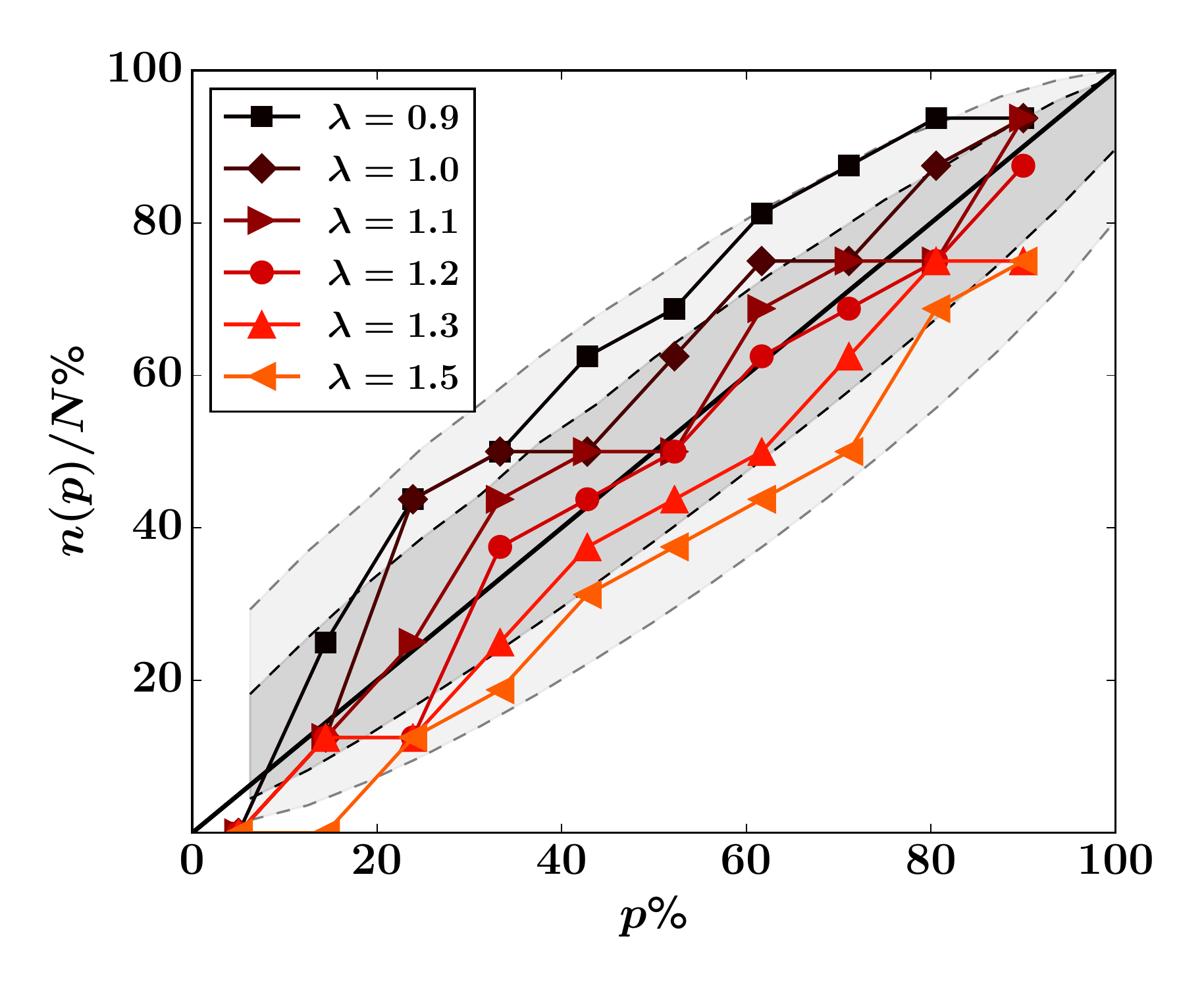}
  \caption{(color online) Empirical determination of $\lambda$
  by comparing results at different orders. The cross sections used are
  the computations with the $R=0.9\,$fm regulator. Priors are Set C$_\epsilon^{(1)}$. For full explanation see text.
  \label{fig:lam-success-plot-C}}
\end{figure}

We also compute the intervals using Set $\Cepsone$, which accounts for the effects
of each coefficient and is less conservative. The results are contained in
Fig.~\ref{fig:lam-success-plot-C}. We see that even for these assumptions,
the $\lambda=1$ curve gets outside the $1\sigma$ band. The plot suggests $\lambda = 1.1$ is a more consistent choice (other values near $\lambda=1.1$ will, of course, also be consistent). 
Because the DOB intervals computed with Set $\Cepsone$ priors are more
informed by the available coefficients, this  result may suggest a small increase in the assigned breakdown scale
is appropriate. However, we note the small amount of data on EFT convergence that is being used here; almost all rescalings considered are consistent at the 2$\sigma$ level. 
Such determinations of $\Lambda_b$ from success rates can be sharpened by considering the behavior of the EFT series for more observables.

\subsection{Gaussian naturalness and the Forte method}

in Ref.~\cite{Forte:2013mda}, Forte et al.\ suggest that, for QCD expansions, the best $\lambda$ is the one that makes all
the expansion coefficients closest to the same size, which they interpret as a
statement that the coefficients should be normally distributed
around a single number $\mu$ with variance $\sigma^2$~\cite{Forte:2013mda}.
For a quantity for which the known information is a mean
and standard deviation,  in
this case a particular coefficient $c_n$,
the method of maximum entropy results in a distribution that is a gaussian~\cite{Gull:98,Sivia:06}:
\beq
    \pr(c_n | \lambda, \mu, \sigma) 
    =  \frac{1}{\sqrt{2 \pi}\sigma} \exp \left(
    -\frac{(|c_n| \lambda^n - \mu)^2}{2\sigma^2} \right) \; .
    \label{eq:cn-naturalness-gaussian}
\eeq
If we have several known coefficients, all of which are drawn from a distribution
with the same mean and standard deviation, the joint pdf
$\pr(c_0, \ldots, c_k|\lambda, \mu, \sigma)$ becomes the standard likelihood function.
If  $\sigma = \cbar$ and
$\mu = 0$ such   a distribution corresponds to the Set C prior of
Table~\ref{tab:priors}.

Forte et al.\
consider the probability distribution for both $\mu$ and $\lambda$ given a
set of $\{c_n\}$~\cite{Forte:2013mda}. This can be obtained from \eqref{eq:cn-naturalness-gaussian} using
Bayes' theorem:
\beq
    \pr(\lambda, \mu | c_0, \ldots, c_k, \sigma) = \frac{\pr(c_0, \ldots, c_k| \lambda, \mu, \sigma)
    \, \pr(\lambda, \mu| \sigma)}{\pr(c_0, \ldots, c_k|\sigma)} \;.
\eeq
Forte et al.\ assign no prior information to $\lambda$ and $\mu$ other than
that both are larger than zero, and neither quantity depends on $\sigma$ \emph{a priori}.
They then take the prior and the evidence in the denominator to be an overall normalization factor
that is independent of $\lambda$ and $\mu$, and so do not calculate them
explicitly (cf.\ discussion of a scale-invariant prior for $\lambda$ below).
 The pdf for $\lambda$ and $\mu$ can then be written
\beq
    \pr(\lambda, \mu | c_0, \ldots, c_k, \sigma) \propto \pr(c_0, \ldots, c_k| \lambda, \mu, \sigma) \;,
    \label{eq:forte-prob-lam-mu}
\eeq
meaning that maximizing the probability of $\lambda$ and $\mu$ is equivalent to
minimizing
\beq
  \chi^2 = \sum_{i=1}^{N_O} \sum_{n=0}^k \left(\frac{|c_n^{(i)} | \lambda^n - \mu}{\sigma}\right)^2 \;,
  \label{eq:chisqForte}
\eeq
where $\{c_n^{(i)}\}$ is the set of EFT coefficients found for the $i$th observable, and 
$N_O$ is the number of observables being used to form the $\chi^2$. In our case $N_O=4$: the 
cross sections at the four different energies analyzed by EKM~\footnote{In general there would be $N_0 n_c$ terms in the $\chi^2$ sum, but we omit the N$^4$LO coefficient from the 50 MeV cross section, since it is clearly an outlier. Our $\chi^2$ thus has 19 terms in the sum.}. Note also that for chiral EFT for NN scattering the coefficient
$c_1$ is known to be zero, and so the $n=1$ term should be omitted from the sum. 

The assumption that
$\lambda$ has a uniform prior is not consistent with arguments regarding the invariance of the pdf under a change of scale~\cite{Jeffreys:1939}. In fact,
$\lambda$ should be treated as a scale parameter. So, in contrast to Ref.~\cite{Forte:2013mda},
we assign a uniform prior to the \emph{logarithm} of $\lambda$, resulting
in a probability distribution for $\lambda$
and $\mu$ that is:
\beq
    \pr(\lambda, \mu | c_0, \ldots, c_k, \sigma) \propto \frac{1}{\lambda} \times e^{-\chi^2/2} \; ,
    \label{eq:mod-prob-lam-mu}
\eeq
with the parameter space for $\lambda$ and $\mu$ restricted to both being positive.
Assuming $\sigma=1$, we find the maximum of the probability \eqref{eq:forte-prob-lam-mu} for 
the $R=0.9\,$~fm EKM coefficients occurs at $\lambda=0.92$, $\mu=0.69$. To consider
the pdf of $\lambda$ only, we marginalize over the parameter $\mu$ and maximize
$\pr(\lambda|c_0,\ldots,c_k,\sigma)$ to find $\lambda = 1.01^{+0.18}_{-0.19}$,
which is consistent with the Forte {\it et al.} hypothesis at a 68\% DOB. 
Larger $\sigma$'s generate still wider ranges. From this point of view too, then,
$\Lambda_b=600$ MeV is a consistent choice for the $R=0.9$ fm np scattering EFT-expansion coefficients.

\subsection{$\chi^2$ test}

Alternatively, we can demand that the mean of the $c_n$'s be fixed at
$\mu = 0$ and that the width $\sigma$ should affect the results as in Set~C
gaussian pdfs on the coefficients, where $\cbar$ is an important feature of
the prior. This leaves us with the probability
\begin{align}
    \pr(\lambda, \mu =0| c_0, \ldots, c_k, \sigma) \propto \frac{1}{\lambda} \exp\left(-\frac{\chi^2(\mu=0)}{2}\right) \; ,
\end{align} 
where $\chi^2(\mu=0)$ is given by Eq.~\eqref{eq:chisqForte} with $\mu=0$. 

We can then test whether, for a given $\lambda$, the data, i.e., the EKM coefficients from their $R=0.9$ fm calculation, follows a normal distribution with mean zero
and width $\sigma$. We do this by comparing $\chi^2(\mu=0)$ with the way that $\chi^2$ should be distributed for 
a normal distribution with $19$ degrees of freedom.%
Once again, in order to do this we must fix $\sigma$. With the choice $\sigma=1$ we find $\lambda=1.09$ 
gives $\chi^2$ of 19---the central value one would expect for this many data points~\footnote{Including the N$^4$LO coefficient from the 50 MeV cross section lowers the
results for $\lambda$ by about 10\%.}. Using the rule of 
thumb for large number of degrees of freedom, $N$,~\cite{NumericalRecipes} that the $\chi^2$ should 
have a width of $\sqrt{2N}$ indicates that $\lambda$ could (68\% DOB) be anywhere 
between 1.01 and 1.15. As in the previous subsection, choices of $\sigma > 1$ will increase this range of possibilities.

\subsection{Summary of expansion-parameter checks}

In any case, while none of these methods provide a crisp result for $\Lambda_b$ from the 19 data points analyzed, it is
reassuring that there is little evidence for a large change in $\Lambda_b$.
Minimally, EKM's estimate $\Lambda_b \approx 600\,$MeV
for their $R=0.9\,$fm calculation is consistent with these analyses, and the breakdown scale may in fact be a little higher.
Further investigations employing these techniques with EFT coefficients drawn from many more observables will provide more definitive answers.

%%%%%%%%%%%%%%%%%%%%%%%%%%%%%%%%%%%%%%%%%%%%%%%%%%%%%%%%%%%%%%

\section{Summary and outlook} \label{sec:conclusion}

We have adapted and extended the Bayesian framework originally introduced in the context of pQCD by Cacciari and
 Houdeau~\cite{Cacciari:2011ze} to evaluate truncation errors in EFT expansions.
Assumptions about the nature of the coefficients in the expansion are encoded as priors
on the coefficients of higher-order terms in the EFT series. The pdfs for these coefficients
then ultimately also include information on the distribution of coefficients at orders
that are calculated. Here we employed priors derived from the notion of
``naturalness" of EFT coefficients, i.e., the idea that they should be $\mathcal{O}(1)$
when the observable and the momentum of the process  in question are measured
in appropriate units. We took the coefficients in the EFT expansion of cross sections
to be natural in this sense. Such a choice is uncontroversial for perturbative processes,
e.g., meson-meson scattering at momenta well below the chiral-symmetry-breaking
scale. It remains to be fully investigated for cross sections in nucleon-nucleon
 scattering, where
the relationship between the underlying scales and observables is quite complex;
we rely here on an empirical validation (see Fig.~\ref{fig:scaledR0p9Lam600}).

We investigated the influence of two prior pdfs for EFT coefficients on the truncation errors. The first was the CH characterization of an upper bound
$\cbar$, the second was a Gaussian of width $\cbar$. We also investigated the influence of
priors on $\cbar$ itself on the results.
We did this in the context of representative examples in Section~\ref{sec:CH} and, in Section~\ref{sec:EKM}, using
results from the order-by-order calculations of neutron-proton cross
sections by Epelbaum, Krebs, and Mei\ss ner (EKM) in Ref.~\cite{Epelbaum:2014sza} (obtained with a regulator parameter
$R=0.9\,$fm).  Combining the insights from both sections we find:
\begin{itemize}
\item  Priors that reflect a natural size for $\cbar$ give similar
degree-of-belief (DOB) intervals at the lowest orders.
\item The resulting error bands are tighter than
those for which the scale of $\cbar$ is not constrained.
\item For higher orders, 68\% DOB intervals show little dependence on prior choice;
 95\% DOB intervals have larger, but still quite small, dependence.
\end{itemize}

These results have wide applicability to observables---they can be used in many EFT contexts. In the case of neutron-proton scattering our formulas provide a statistical interpretation to
error bars obtained by EKM in Ref.~\cite{Epelbaum:2014efa}. Their error bar is obtained in the case that the distribution of coefficients is uniform, in which case it
is a $j/(j+1)*100\%$ DOB interval for the omitted terms in a N$^j$LO calculation. But, as already stated, truncation errors in these calculations at
NLO or beyond (i.e., which include  at least two orders beyond leading) were only mildly dependent on
prior choice. In particular, the 68\% DOB intervals obtained in our Bayesian framework varied by at most 15\% amongst all the priors considered here, and the variation was less than that
in calculations beyond NLO. Error bands at a given order were also consistent with a statistical interpretation when compared with known higher-order results.
Truncation errors at leading order {\it are} sensitive to
prior choice, since--given the choice of scaling observable we made---almost no information on the pattern of coefficients emerges from a leading-order calculation. Comparison of the resulting error band with
the known results of NLO, N$^2$LO, N$^3$LO, and N$^4$LO calculations suggests that the CH choice of a $\theta$-function distribution for coefficients, and a scale-independent
distribution for the width of that $\theta$-function, is too conservative---at least for this case. Overall then, at sufficiently high order, the prior picked from Table~\ref{tab:priors} hardly matters---in practice $k=2$ may be enough. At lower orders priors provide a rigorous way to explore different assumptions about the pattern of
coefficients in the EFT.

Indeed, the application of Bayesian methods to data is often criticized because
of the apparently subjective selection of prior pdfs. However, the
priors manifest what would otherwise be implicit assumptions, so that they
can be tested. The information encoded in those assumptions is then modified
in light of subsequent data: in this case the distribution of low-order
coefficients influences the distributions computed for coefficients
that enter the assessment of the truncation error.  Furthermore, the development
of specific pdfs for those higher-order coefficients allows a statistical interpretation
of the ``theory error"---or at least the part of it that results from the truncation of
the EFT series. This allows crisp answers to questions regarding, for example,
how theory error bars should be combined, or the extent to which theory errors
on different quantities are correlated. Those answers may have some sensitivity
to the choice of prior on the higher-order coefficients, but the advantage of the
Bayesian framework is that the consequences of prior assumptions about the distribution
of coefficients (Uniformly distributed or Gaussian? Natural or Scale-less?) can be traced
through to the statistical uncertainties on the EFT calculation. Those assumptions can then---if
necessary---be refined.

Such refinement may be necessary in light of the need to identify an EFT breakdown scale before extracting the (supposedly) $\mathcal{O}(1)$ coefficients which are input to our analysis.
Mis-identification of the breakdown scale is one manner in which a particular prior could fail. But, in this case, we showed in Sec.~\ref{sec:scalechecking} that
this breakdown scale $\Lambda_b=600$ MeV leads to success rates taken from the EFT predictions at four different energies, and for four different orders, that are statistically consistent with the DOB intervals resulting from our Bayesian formalism.
Furthermore, the distribution of coefficients with the $R=0.9$ fm regulator choice is consistent with a Gaussian distribution. Qualitatively  a natural distribution is not seen for the coefficients obtained
using a second, larger, value of $R$. The calculation at this larger regulator radius reflects cutoff artifacts,
which leads to peculiar convergence of the EFT expansion. The breakdown of the EFT is then not set by $\Lambda_b$, but by the effects of this softer cutoff.
With the general formalism for probability distributions of EFT coefficients laid out here
it will be important to check when the EFT coefficients obtained over
a wide range of cutoff values and
 observables are empirically consistent with the application of naturalness priors
 to observables in the NN system.

The Bayesian approach to error estimation presented here is an alternative
to procedures that calculate error bands based on variation of the
EFT regulator, which could be a cutoff in either momentum or coordinate space.
While variation with regulator scale gives a lower bound on the uncertainty (theories should,
after all, be regulator invariant up to higher-order terms) the resulting error band has
no statistical interpretation.
A particular flaw is the arbitrariness of the interval in which the cutoff is varied;
for QCD this is only of mild concern, because the dependence on the regulator parameter is
only logarithmic. But running in the chiral EFT applied to NN scattering is much faster: it can contain
high positive powers of the regulator (momentum) scale. This concern is
exacerbated by the narrow range that is possible before encountering
 irremediable cutoff artifacts or spurious deep bound states.
It is also the case that residual cutoff dependence only reflects the
contribution from omitted contact operators. These only enter the chiral expansion for NN observables
at even orders, and so examination of cutoff dependence alone may substantially
underestimate the EFT truncation error.
More generally, when computed using only cutoff variation, the error bands for predictions
of observables (as opposed to quantities used to fit EFT LECs) generically exhibit undesirable
systematics (e.g., sometimes growing wider with order) and often underestimate the
error when compared with actual higher-order calculations~\cite{Furnstahl:2014jpg}.
In contrast, the Bayesian assessment of truncation errors laid out here is applicable to all EFTs,  admits a statistical interpretation
of truncation errors,
is justified when regulator parameters cannot be
varied widely, and predicts decreased
errors at \emph{all} orders---not just when new LECs are added.

The truncation-error assessment described here is just one piece of a broader framework for
EFT uncertainty quantification using Bayesian methods.
We have under development analogous procedures, together with a suite of diagnostic
tools, for parameter estimation and the assessment and propagation of errors---both statistical and truncation---in
fitted LECs and predicted observables.
Bayesian model selection is also well suited for addressing fundamental
questions in nuclear EFT, such as the comparative efficacy of theories with
different degrees of freedom, from pionless to chiral EFTs with and without
an explicit $\Delta(1232)$.

\acknowledgments{We are grateful to Evgeny Epelbaum for numerous discussions on these issues, and for sharing
results prior to publication. We also thank Harald Grie\ss hammer for a careful reading of, and useful suggestions regarding,
an early version of this manuscript. NMK thanks the KITP (Klco Institute for Theoretical Physics) for hospitality during the
completion of this research.
This work was supported in part by the National Science Foundation
under Grant No.~PHY--1306250, the U.S. Department of Energy under
grant DE-FG02-93ER40756, and the NUCLEI SciDAC Collaboration under
DOE Grant DE-SC0008533.}

%%%%%%%%%%%%%%%%%%%%%%%%%%%%%%%%%%%%%%%%%%%%%%%%%%%%%%%%%%%%%%

%\bibliography{bayesian_refs}

\end{document}